\documentclass[12pt, draftclsnofoot, onecolumn]{IEEEtran}
\usepackage[utf8]{inputenc}
\usepackage{bbm,dsfont,mathrsfs,amsthm}
\usepackage{bbm,dsfont,mathrsfs,fixmath,amsthm}
\usepackage{amsmath,amssymb,graphicx} 
\usepackage{amssymb}
\usepackage{graphicx}
\usepackage{cite}
\usepackage{comment}
\usepackage{epsfig}
\usepackage{color}
\usepackage{graphicx}
\usepackage[nolist,printonlyused]{acronym}
\usepackage{cancel}
\usepackage{ctable}
\newcommand{\gTX}{g_{\mathrm{tx}}}
\newcommand{\gRX}{g_{\mathrm{rx}}}
\newcommand{\Xnm}{X\left[n,m\right]}
\newcommand{\XnmPrime}{X\left[n',m'\right]}
\newcommand{\HnmPrime}{H_{n,m}\left[n',m'\right]}
\newcommand{\nPrimeSum}{\sum_{n'=0}^{N-1}}
\newcommand{\mPrimeSum}{\sum_{m'=0}^{M-1}}
\newcommand{\kPrimeSum}{\sum_{k'=0}^{N-1}}
\newcommand{\lPrimeSum}{\sum_{l'=0}^{M-1}}

\newcommand{\pSum}{\sum_{p=0}^{P-1}}
\newcommand{\PsiMat}{\boldsymbol{\Psi}^p_{k,k'}\left[l,l'\right]}
\newcommand{\Pavg}{P_{\mathrm{avg}}}
\newcommand{\rect}{{\rm rect}}

\newcommand{\Id}{{\bf I}}

\begin{document}

	\begin{acronym}
		\acro{AWGN}{additive white Gaussian noise}
		\acro{OTFS}{orthogonal time frequency space}
		\acro{ISFFT}{inverse symplectic finite Fourier transform}
		\acro{SFFT}{symplectic finite Fourier transform}
		\acro{CRLB}{Cram\'er-Rao lower bound}
		\acro{ISI}{inter-symbol interference}
		\acro{ICI}{inter-carrier interference}
		\acro{SNR}{signal-to-noise ratio}
		\acro{SINR}{signal-to-interference noise ratio}
		\acro{MAE}{mean absolute error}
		\acro{OFDM}{orthogonal frequency division multiplexing}
		\acro{CP}{cyclic prefix}
		\acro{DFT}{discrete Fourier transform}
		\acro{IDFT}{inverse discrete Fourier transform}
		\acro{ML}{Maximum Likelihood}
		\acro{MMSE}{minimum mean square error}
		\acro{FMCW}{frequency modulated continuous wave}
		\acro{MSE}{mean square error}
		\acro{CSI}{channel state information}
		\acro{MP}{message-passing}
		\acro{LLR}{log-likelihood ratio}
		\acro{MFUB}{matched filter upper bound}
		\acro{BPSK}{binary phase-shift keying}
		\acro{QAM}{quadrature amplitude modulation}
		\acro{BER}{bit error rate}
		\acro{PL}{path loss}
		\acro{UB}{upper bound}
		\acro{LoS}{line-of-sight}
		\acro{FG}{Factor Graph}
		\acro{pdf}{probability density function}
		\acro{SPA}{sum-product algorithm}
		\acro{MIMO}{multiple-input multiple-output}
		\acro{LMMSE}{linear minimum mean square error}
		\acro{LDPC}{low-density parity check}
	\end{acronym}
	
	%
	\title{On the Effectiveness of OTFS for Joint Radar and Communication}

	\author{
		\IEEEauthorblockN{Lorenzo Gaudio$^{1}$, Mari Kobayashi$^{2}$, Giuseppe Caire$^{3}$, and Giulio Colavolpe$^{1}$}\\
		\IEEEauthorblockA{$^{1}$University of Parma, Italy \\
			$^{2}$Technical University of Munich, Munich, Germany \\
			$^{3}$Technical University of Berlin, Germany
			\thanks{Emails: lorenzo.gaudio@studenti.unipr.it, mari.kobayashi@tum.de,~caire@tu-berlin.de, giulio.colavolpe@unipr.it}
			}}
	
	\maketitle
	
	\vspace{-1.5cm}
	
	\begin{abstract}
		We consider a joint radar estimation and communication system using orthogonal time frequency space (OTFS) modulation. The scenario is motivated by vehicular applications where a vehicle equipped with a mono-static radar wishes to communicate data to its target receiver, while estimating parameters of interest related to this receiver. In a point-to-point communication setting over multi-path time-frequency selective channels, we study the joint radar and communication system from two perspectives, i.e., the radar estimation at the transmitter as well as the symbol detection at the receiver. For the radar estimation part, we derive an efficient approximated Maximum Likelihood algorithm and the corresponding Cram\'er-Rao lower bound for range and velocity estimation. Numerical examples demonstrate that multi-carrier digital formats such as OTFS can achieve as accurate radar estimation as state-of-the-art radar waveforms such as frequency-modulated continuous wave (FMCW). For the data detection part, we focus on separate detection and decoding and consider a soft-output detector that exploits efficiently the channel sparsity in the Doppler-delay domain. We quantify the detector performance in terms of its {\em pragmatic capacity}, i.e. the achievable rate of the channel induced by the signal constellation and the detector soft output. Simulations show that the proposed scheme outperforms concurrent state-of-the-art solutions. Overall, our work shows that a suitable digitally modulated waveform enables to efficiently operate joint radar and communication by achieving full information rate of the modulation and near-optimal radar estimation performance. Furthermore,  OTFS appears to be particularly suited to the scope. 
	\end{abstract}
	
	
	\section{Introduction}  \label{intro}
	
	A key-enabler of high-mobility networks is the ability of a node to continuously track its dynamically changing environment (state) 
	and react accordingly. 
	Although state sensing and communication have been designed separately in the past, 
	power and spectral efficiency and hardware costs encourage the integration of these two functions, such that they are 
	operated by sharing the same frequency band and hardware (see e.g. \cite{zheng2019radar}). 
	Motivated by emerging vehicular applications (V2X) \cite{chen2017vehicle}, we consider a {\it joint radar and communication} 
	system where a radar-equipped transmitter wishes to transmit information to a target receiver and simultaneously 
	estimate the parameters of that receiver such as range and velocity. 
	
	Such a communication setup has been extensively studied in the literature (see \cite{sturm2011waveform,patole2017automotive, kumari2018ieee, nguyen2017delay,ma2019joint,dokhanchi2019adaptive,mishra2019towards,dokhanchi2019mmwave,hassanien2019dual} and references therein). 
	Existing works on joint radar and communications can be roughly classified into two classes. 
	The first class considers some resource-sharing approach, such that time, frequency, or space resources are split into either radar or data communication 
	(e.g., see \cite[Section III. A, C]{ma2019joint} and references therein). 
	The second class uses a common waveform for both radar and communication. This approach includes information-embedded radar waveforms (e.g., see \cite[Section III. D]{ma2019joint}, \cite[Section IV. A]{zheng2019radar}, \cite{dokhanchi2019mmwave}, and references therein) as well as 
	the direct usage of standard communication waveforms applied to radar detection (e.g., see \cite[Section IV. B]{zheng2019radar}, \cite[Section III. B]{zheng2019radar}, \cite{sturm2011waveform, braun2014ofdm,liu2017adaptive,nguyen2017delay,viterboOTFSradar}). 
	It is worth noticing that a synergistic waveform design for joint radar and communications yields a significant 
	potential gain compared to the resource-sharing approach, as demonstrated in a simplified albeit representative 
	information theoretic framework in \cite{kobayashi2018joint}. 
	It is also worthwhile to notice that typical V2X channels are characterized by a relatively small number of discrete multi-path components corresponding to \ac{LoS} propagation, ground reflection, and some specular reflections on surrounding (e.g., metal) surfaces, each of which is characterized by its own, possibly large, Doppler frequency shift (e.g., see \cite{mahler2016measurement,mahler2016tracking} and references therein). Motivated by this fact, we focus on a joint radar and communication system using \ac{OTFS} modulation (see \cite{hadani2017otfs,raviteja2018interference} and references therein), 
	as this modulation format provides inherent robustness to Doppler shifts and is naturally suited to sparse channels in the delay-Doppler domain \cite{shen2019channel,raviteja2019OTFSchannelEst}.
		
	The first part of this paper focuses on the suitability of \ac{OTFS} for joint radar parameter estimation and data communications, by considering 
	the \ac{MSE} of range and velocity estimation and the achievable rate with Gaussian inputs 
	for a simple point-to-point communication scenario. 
	We extend our preliminary work \cite{gaudio2019performance}, which considered only \ac{LoS} propagation, to  the case of a channel with multiple paths, one of which is the \ac{LoS}.  We propose an efficient approximated \ac{ML} algorithm to estimate the range and velocity of the target from the backscattered signal and derive the corresponding \ac{CRLB}.
	Our numerical examples inspired by the parameters of IEEE 802.11p demonstrate that digital multi-carrier modulation formats 
	such as \ac{OTFS} and \ac{OFDM} yield as accurate estimation performance as  \ac{FMCW}, 
	one of the typical automotive radar waveforms \cite{patole2017automotive}, while achieving significant communication rates. 
	These results suggest that joint radar and communication can be effectively operated without compromising neither the achievable data rates 
	nor the radar performance.
		
	The second part of the paper addresses more specifically the soft-output 
	symbol detection for \ac{OTFS} at the receiver side. We consider separate detection and decoding and evaluate several options for 
	soft-output symbol detection in terms of their {\em pragmatic capacity}, i.e., 
	the achievable rate of the channel induced by the signal constellation used with uniform probability at the input and the detector soft 
	output \cite{kavcic2003binary,soriaga2007determining}. 
	 This performance metric characterizes information theoretical rates achievable by separate detection and decoding, when a specific signal constellation 
	 and a specific soft-output symbol detector are employed. Therefore, it is more meaningful and practically relevant than uncoded bit error 
	 rate, as usually considered in concurrent works.	
	We propose a \ac{MP} detector derived from the general approach of \cite{colavolpe2011SISOdetection}, 
	suitably adapted to the \ac{OTFS} signal format. The proposed scheme is compared with the following schemes: i) another MP-based scheme recently proposed in 
	\cite{raviteja2018interference}; ii) the standard linear \ac{MMSE} block equalizer, which is prohibitively complex due to its computation of a large-dimensional matrix inversion; iii) a recently proposed low-complexity approximated linear \ac{MMSE} equalizer \cite{cheng2019low}, based on some drastic 
	simplifying assumptions that provide a conveniently structured channel matrix for which the large dimensional matrix inversion can be avoided. 
 Our simulation results show that the proposed detection scheme significantly outperforms all other low-complexity schemes, and it is also able to 
outperform the high-complexity \ac{MMSE} block equalizer when the channel is sufficiently sparse. 

	The main contributions of this work are summarized as follows: \\
	1) we propose an efficient algorithm to estimate range and velocity for \ac{OTFS} modulation employing practical rectangular pulses over a $P$-path time-frequency selective radar channel. The proposed algorithm is different from the one recently proposed in \cite{viterboOTFSradar}. 
	While  \cite{viterboOTFSradar} focuses on a low-complexity matched filter approach, we consider \ac{ML} parameter estimation. 
	Our proposed algorithm coincides with the \ac{ML} estimator for a single path channel ($P=1$) and it performs also very well for $P > 1$. 
	Furthermore, our proposed algorithm is able to handle continuous values of delay and Doppler, contrary to  \cite{viterboOTFSradar} that assumes
	discretized delays. \\
	2) We derive the \ac{CRLB} of range and velocity estimation for a general $P$-path time-frequency selective channel. 
	For the special case of a single-path channel ($P=1$), we also analyze the typical transition behavior of the \ac{ML} estimator (usually referred to as 
	{\it waterfall}), following an approach similar to \cite{athley2003space,athley2005threshold}. \\
	3) Through these analytical results and of computer simulations, we show that fully digitally modulated waveforms such as\ac{OTFS} and \ac{OFDM} provide as accurate range and velocity estimation as \ac{FMCW}, one of typical radar waveforms. Furthermore, \ac{OTFS} achieves slightly better data rates than \ac{OFDM} since the latter incurs a higher overhead due to the \ac{CP}. \\
	4) We adapt the \ac{MP}-based detection scheme of \cite{colavolpe2011SISOdetection} to the \ac{OTFS} format, evaluating its performance in terms of pragmatic capacity. Our scheme improves upon existing works in various aspects. 
	First,  we consider \ac{OTFS} modulation with practical rectangular pulses, rather than ideal pulses satisfying the bi-orthogonal condition as considered in \cite{hadani2017otfs,cheng2019low,nimr2018extendedGFDM}. Unfortunately such ideal pulses are mathematically impossible to construct \cite{matz2013time}. Second, our method constructs a {\em \ac{FG}} (e.g. \cite{Kschischang2001SPA}) of girth 6, for which the exact \ac{SPA} (e.g. \cite{Kschischang2001SPA}) can be directly applied with linear complexity in the symbol constellation size, as opposed to the \ac{MP} detector of \cite{raviteja2018interference}, based  on a \ac{FG} of girth 4, and requiring a Gaussian approximation of the interfering constellation symbols in order to avoid high complexity computations (more details on this point are discussed in  Section \ref{sec:Detection-Algorithm}). Finally, our scheme does not make simplifying assumptions on the channel model such as discretized delay and Doppler shifts and/or 
	exactly bi-orthogonal pulses, therefore it is much more robust to realistic system and channel conditions.
		
	The paper is organized as follows. In Section \ref{sec:phy-model}, we present the physical model. In Section \ref{sec:OTFS-channel}, we derive the \ac{OTFS} input-output relation and the definition of the channel matrix. In Section \ref{sec:OTFS-radar-estimation}, we provide the \ac{ML} radar estimator together with the theoretical analysis of the \ac{ML} estimation performance. Section \ref{sec:Detection-Algorithm} introduces the considered symbol detection algorithms. 
	In Section \ref{sec:sim-results}, we define  the chosen performance metrics and then we show some illustrative numerical results. 
	Finally, Section \ref{sec:conclusions} concludes the paper. 
	
	\section{Physical model}\label{sec:phy-model}

	We consider a joint radar and communication system over a channel bandwidth $B$ operating at the carrier frequency $f_c$. We assume that a transmitter, equipped with a mono-static full-duplex radar, wishes to convey a message to its target receiver while estimating parameters of interest related to the same receiver from the backscattered signal. 
	Full-duplex operations can be achieved with sufficient isolation between the transmitter and the (radar) detector and possibly interference analog pre-cancellation in order to prevent the (radar) detector saturation \cite{sabharwal2014band}. 
	For simplicity, in this paper we consider no self-interference, keeping in mind that some residual self-interference can be handle as additional noise \cite{braun2014ofdm}. 
	
	We model the channel as a $P$-taps time-frequency selective channel given by 
	\begin{align} \label{eq:h_channel}
	h(t, \tau) = \sum_{p=0}^{P-1}{h_p} e^{j2\pi \nu_p t} \delta\left(\tau-\tau_p\right)\,,
	\end{align}
	where $P$ denotes the number of scattering reflections or paths and the 0-th path is \ac{LoS}, $h_p$ is a complex channel gain including the \ac{PL} of the path component, while $\nu_p = \frac{v_p f_c}{c}$, $\tau_p=\frac{r_p}{c}$, $v_p$, and $r_p$ denote the corresponding one-way Doppler shift, delay, velocity, and range associated to the $p$-th propagation path, respectively. The round-trip Doppler shift and delay at the radar detector (co-located with the transmitter) are obtained by doubling the one-way values of the \ac{LoS} path. 
	
\begin{figure}
		\centering
		\includegraphics[scale=0.65]{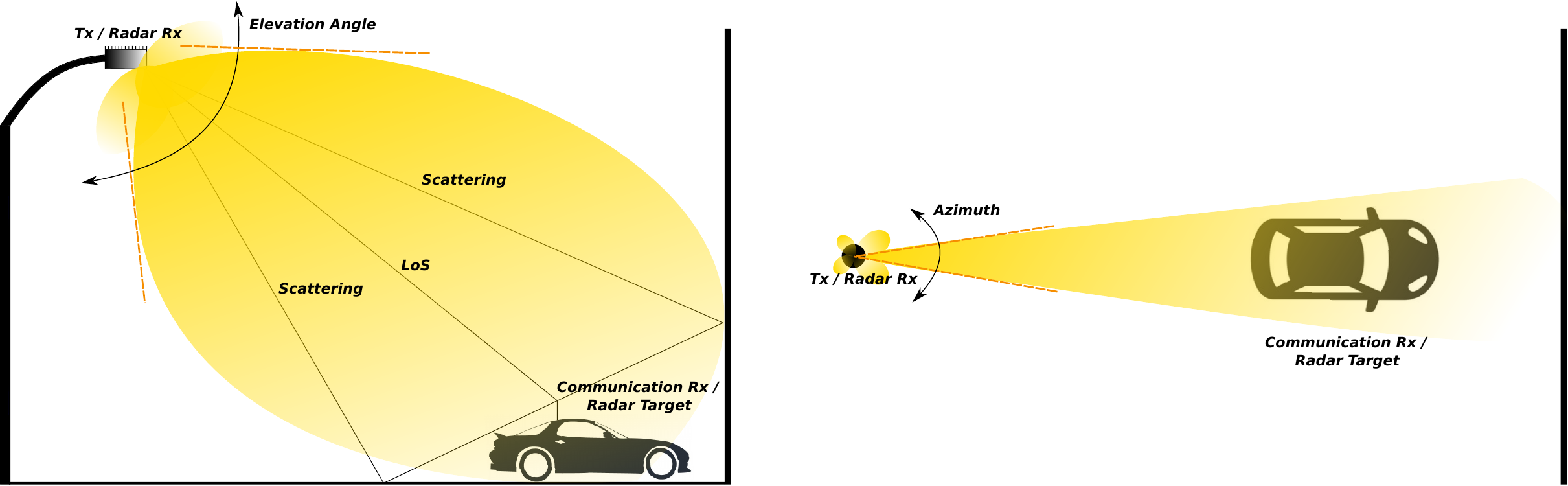}
		\caption{The figure shows the beam sweeping scenario of our work. Between the transmitter (Tx) and the radar target there is a LoS component and some scattering paths, depending on the surrounding environment. The Rx (e.g., a car in this case) ``passively'' 
		reflects the incoming signal back to the Radar Rx (co-located with the Tx).}
		\label{fig:radar-comm-scenario}
	\end{figure}

	We assume that the transmitter periodically scans narrow angular sectors (e.g., using a phased array). 
	If the sector beam is narrow enough, it is reasonable to consider a single target in each sector, while the propagation 
	can go through multiple paths, one of which is the \ac{LoS}, and the others may be caused by ground reflection and reflections on buildings and metal surfaces (e.g., vehicles parked along the street). An example of this scenario is depicted in Fig. \ref{fig:radar-comm-scenario}. Therefore, we focus on the estimation of the delay (related to the range between transmitter and receiver) and of the Doppler shift (related to the relative velocity between transmitter and receiver) of the \ac{LoS} path, since in the envisaged scenario the direction of arrival is automatically determined by the scanned angular sector. 
	
	\section{OTFS Input-Output Relation}\label{sec:OTFS-channel}
	
	We derive here the input-output discrete complex baseband model for the \ac{OTFS} signal format by generalizing previous works \cite{hadani2017otfs,raviteja2018interference}, removing certain oversimplifying assumptions such as bi-orthogonal pulses and discretized delay and/or Doppler shifts. 
	
	As usually done in \ac{OTFS} modulation (e.g., \cite{hadani2017otfs,raviteja2018interference}), data symbols $\left\{x_{k,l}\right\}$, for $k=0,\dots,N-1$ and $l=0,\dots,M-1$, are arranged in an $N\times M$ two-dimensional grid referred to as the Doppler-delay domain. 
	We consider the average power constraint given by 
	\begin{equation}\label{eq:avg-power}
	\frac{1}{NM}\sum_{k=0}^{N-1}\sum_{l=0}^{M-1}\mathbb{E}\left[\left|x_{k,l}\right|^2\right]\leq\Pavg\,.
	\end{equation}
	
	In order to send the block of symbols $\{x_{k,l}\}$, the transmitter first applies the \ac{ISFFT} to convert data symbols into a block of 
	samples $\{X[n, m]\}$ in the dual domain, referred to as the time-frequency domain. This is given by 
	\begin{equation}\label{eq:x-to-X}
	\Xnm=\sum_{k=0}^{N-1}\sum_{l=0}^{M-1}x_{k,l}e^{j2\pi\left(\frac{nk}{N}-\frac{ml}{M}\right)}, 
	\end{equation}
	for $n = 0, \ldots, N-1$ and $m = 0, \ldots, M-1$. 
	Then, it generates the continuous-time signal
	\begin{equation}
	s\left(t\right)=\hspace{-0.1cm}\sum_{n=0}^{N-1}\sum_{m=0}^{M-1}\Xnm\gTX\left(t-nT\right)e^{j2\pi m\Delta f\left(t-nT\right)},\,\hspace{-0.18cm}
	\end{equation}
	where $\gTX\left(t\right)$ denotes the transmit shaping pulse, $T$ is the symbol duration, 
	and  we consider a multi-carrier system where the total bandwidth is divided into $M$ subcarriers, i.e., $B=M \Delta f$, where $\Delta f = 1/T$ denotes the subcarrier spacing. By considering $\gTX\left(t\right)$ strictly time limited in $\left[0,T\right]$, the \ac{OTFS} frame duration is given by  
	$T_f^{\mathrm{otfs}}=NT$ (generally the duration is larger, but in practice it is well approximated by $NT$ up to some guard interval 
	containing the ``tail'' of $\gTX\left(t\right)$). The noiseless received signal $r\left(t\right)$ after transmission through the time-frequency selective channel in \eqref{eq:h_channel} is given by 
	\begin{equation}\label{eq:OTFS-RxSignal}
	r(t) = \int h(t,  \tau) s(t-\tau) d\tau = \sum_{p=0}^{P-1} h_ps(t-\tau_p)e^{j2\pi \nu_p t}\,,
	\end{equation}
	while the output of the receiver filter-bank adopting a generic receive shaping pulse $\gRX\left(t\right)$ is
	\begin{equation}
	Y\left(t,f\right)
	=\int
	r\left(t'\right)\gRX^*\left(t'-t\right)e^{-j2\pi ft'}dt'\,.   \label{ziocanale}
	\end{equation}
	By sampling at $t=nT$ and $f=m\Delta f$, the received samples in the time-frequency domain are given by 
	\begin{align}\label{eq:Y-time-frequency}
	Y\left[n,m\right]=Y\left(t,f\right)|_{t=nT,f=m\Delta f}=\nPrimeSum\mPrimeSum\XnmPrime\HnmPrime\,,
	\end{align}
	where, by letting $h_p'\triangleq h_pe^{j2\pi\nu_p\tau_p}$, we have
	\begin{align}\label{eq:363}
	H_{n,m}\left[n',m'\right]\triangleq\pSum h_p'e^{j2\pi n'T\nu_p}e^{-j2\pi m\Delta f\tau_p} C_{\gTX,\gRX}\left(\left(n-n'\right)T-\tau_p,\left(m-m'\right)\Delta f-\nu_p\right)\,,
	\end{align}
	where  $C_{u,v}\left(\tau,\nu\right)\triangleq\int_{-\infty}^{\infty} u\left(t\right)v^*\left(t-\tau\right)e^{-j2\pi\nu t}dt$ denotes
	the cross-ambiguity function between two generic pulses $u$ and $v$, as defined in \cite{matz2013time}.
	Finally, the received samples in the Doppler-delay domain are obtained by applying the \ac{SFFT} to \eqref{eq:Y-time-frequency}, i.e.,
	\begin{align}\label{eq:y-sampled}
	y\left[k,l\right]=\frac{1}{NM}\sum_{n=0}^{N-1}\sum_{m=0}^{M-1} Y\left[n,m\right]e^{-j2\pi\left(\frac{nk}{N}-\frac{ml}{M}\right)}=\kPrimeSum\lPrimeSum x_{k',l'} g_{k,k'}\left[l,l'\right]\,,
	\end{align}
	where the \ac{ISI} coefficient of the Doppler-delay pair $\left[k',l'\right]$ seen by sample $\left[k,l\right]$ is given by
	\begin{equation}
	g_{k,k'}\left[l,l'\right]=\pSum h_p'\PsiMat\,,
	\end{equation}
	where the channel matrix entries $\PsiMat$ are defined in \eqref{eq:Psi}.
	\begin{figure*}
		\begin{align}\label{eq:Psi}
		\PsiMat=&\hspace{-0.1cm}\sum_{n,n',m,m'}\hspace{-0.1cm}e^{j2\pi n'T\nu_p}e^{-j2\pi m\Delta f\tau_p}e^{-j2\pi\left(\frac{nk}{N}-\frac{ml}{M}-\frac{n'k'}{N}+\frac{m'l'}{M}\right)}\nonumber\\
		&\times\frac{C_{\gRX,\gTX}\left(\left(n-n'\right)T-\tau_p,\left(m-m'\right)\Delta f+\nu_p\right)}{NM}\,.
		\end{align}
		\noindent\rule{\textwidth}{0.4pt}
	\end{figure*}
	Writing the $N\times M$ matrices of transmitted symbols and received samples as $NM$-dimensional column vectors (stacking the columns of the corresponding matrices on top of each other), we obtain the block-wise  input-output relation as
	\begin{equation}\label{eq:y}
	\boldsymbol{y} = \underbrace{\left(\pSum h_p'\boldsymbol{\Psi}_p\right)}_{\boldsymbol{\Psi}} \boldsymbol{x}+\boldsymbol{w}\,,
	\end{equation}
	where $\boldsymbol{\Psi}_p$ is the $NM\times NM$ matrix obtained from \eqref{eq:Psi} while $\boldsymbol{w}$ denotes the \ac{AWGN} with 
	zero mean and covariance $\sigma_w^2 \Id_{NM}$. Notice that our input-output relation in \eqref{eq:y} is exact (i.e., no approximation was made in its derivation) and holds for any pair of transmit/receive pulses. 
	
	In order to proceed further in a tractable manner, as done in \cite{raviteja2018interference}, 
	we approximate the integral of the cross-ambiguity function with a discrete sum. In particular, we write
	\begin{align}\label{eq:cross-ambiguity-general}
	C_{\gRX,\gTX}\left(\tau,\nu\right)&=\int_0^T\gTX\left(t\right)\gRX^*\left(t-\tau\right)e^{-j2\pi\nu t}dt\nonumber\\
	&\approx\frac{T}{M}\sum_{i=0}^{M-1}\gTX\left(i\frac{T}{M}\right)\gRX^*\left(i\frac{T}{M}-\tau\right)e^{-j2\pi\nu i\frac{T}{M}}\,,
	\end{align}
	where, for analytical convenience, the number of equally spaced discretization nodes is equal to the 
	number of subcarriers $M$. This approximation is accurate provided that $M$ large enough, which is the typical case. 
	Now, by letting $\gTX\left(t\right)$ and $\gRX\left(t\right)$ be rectangular pulses of length $T$, by limiting the channel delay to $\tau_{\mathrm{max}}<T$, it readily follows that the cross-ambiguity function yields non-zero samples only for $n'=n$ and $n'=n-1$.
	\begin{figure*}
		\begin{align}\label{eq:Psi-final}
		\PsiMat
		\approx& \frac{1}{NM}\frac{1-e^{j2\pi\left(k'-k+\nu_pNT\right)}}{1-e^{j2\pi\frac{\left(k'-k+\nu_pNT\right)}{N}}}\frac{1-e^{j2\pi\left(l'-l+\tau_pM\Delta f\right)}}{1-e^{j2\pi\frac{\left(l'-l+\tau_pM\Delta f\right)}{M}}}e^{j2\pi\nu_p\frac{l'}{M\Delta f}}\nonumber\\
		&\times\begin{cases}
		\begin{array}{ll}
		1 & \mbox{if} \;\; l'\in{\cal L}_{\mathrm{ICI}} \\
		e^{-j2\pi\left(\frac{k'}{N}+\nu_pT\right)} & \mbox{if} \;\; l' \in {\cal L}_{\mathrm{ISI}} \\
		\end{array}\hspace{-0.13cm}.
		\end{cases}\hspace{-0.87cm}
		\end{align}
		\noindent\rule{\textwidth}{0.4pt}
	\end{figure*}
	Hence, the entries of the cross-talk matrix $\PsiMat$ can be thus written as in \eqref{eq:Psi-final}, in which we exploited the expression of the approximated cross-ambiguity function for rectangular pulses given by
	\begin{align}\label{eq:cross-ambiguity}
	C_{\gRX,\gTX}\left(\tau,\nu\right)\approx \frac{1}{M}  \sum_{i=0}^{M-1-l_{\tau}} \exp\left(j 2\pi \nu i\frac{T}{M}\right)\,,
	\end{align}
	with $l_\tau\triangleq\lceil\tau/\left(T/M\right)\rceil$, and where we
	defined the index sets
	\begin{equation}
	\begin{cases}
	{\cal L}_{\mathrm{ICI}}\triangleq \left[0,M-1-l_{\tau_p}\right]\\
	{\cal L}_{\mathrm{ISI}}\triangleq \left[M-l_{\tau_p},M-1\right]
	\end{cases}.
	\end{equation} 
	
	Moreover, in order to have the received signal bandwidth still approximately equal to $B$, we must assume that the bandwidth expansion due to Doppler is negligible with respect to $B$. In our case, for the sake of simplicity, we make the usual assumption that $\nu_{\max} < \Delta f$.\footnote{Note that this approximation can be justified in a number of scenarios. For example, consider a scenario inspired by IEEE 802.11p with $f_c=5.89$ GHz and the subcarrier spacing $\Delta f=156.25$ KHz. This yields $v_{\max} \ll 14325$ km/h, which is reasonable even for a relative speed of 400 km/h. The same holds for IEEE 802.11ad with $f_c$ = 60 GHz and $\Delta f = 5.15625$ MHz \cite{cordeiro2010ieee}.}

	\section{OTFS Radar Estimation}\label{sec:OTFS-radar-estimation}
	
	\subsection{Maximum Likelihood Estimator}
	
		Focusing on the channel model defined in \eqref{eq:h_channel} with a single target and multiple backscattered paths, 
	we wish to find the \ac{ML} estimator for the set of $3P$ unknown parameters $\bar{\boldsymbol{\theta}}=\left(\bar{h}_0',\dots,\bar{h}'_{P-1},\bar{\tau}_0,\dots,\bar{\tau}_{P-1},\bar{\nu}_0,\dots,\bar{\nu}_{P-1}\right)$, where the bar indicates the true value of the unknown parameters. 

	By using expression \eqref{eq:Psi-final} for the channel matrix coefficients, which contains the dependency on the parameters $\{\tau_p,\nu_p\}$, 
	the log-likelihood function to be minimized is given by  
	\begin{equation}\label{eq:likelihood-OTFS}
	l\left(\boldsymbol{y}|\boldsymbol{\theta},\boldsymbol{x}\right)=\left\|\boldsymbol{y}-\sum_{p=0}^{P-1}h_p'\boldsymbol{\Psi}_p\boldsymbol{x}\right\|^2\,,
	\end{equation}
	where symbols in $\boldsymbol{x}$ are known at the radar detector, since it is co-located with the transmitter. 
	The \ac{ML} estimator is given by 
	\begin{equation}\label{eq:OTFS-arg-max-ML}
	\hat{\boldsymbol{\theta}}=\arg\min_{\boldsymbol{\theta}\in\mathbb{C}^P\times\mathbb{R}^P\times\mathbb{R}^P}l\left(\boldsymbol{y}|\boldsymbol{\theta},\boldsymbol{x}\right)\,.
	\end{equation}
	A brute-force search for the maximum in a $3P$ dimensional continuous domain is infeasible in general. 
	Therefore, in the following we propose a viable method to approximate the \ac{ML} solution with low complexity.
	
	The log-likelihood function in \eqref{eq:likelihood-OTFS} is quadratic in the complex amplitudes $\{h'_p\}$ for given $\{\tau_p,\nu_p\}$. Hence, 
	 the minimization of \eqref{eq:likelihood-OTFS}  with respect to $\{h'_p\}$ for fixed $\{\tau_p,\nu_p\}$ is readily given as the solution of 
	 the linear system of equations	
	\begin{equation}\label{eq:l-equality}
	\begin{aligned}
	\sum_{q=0}^{P-1}h'_q \boldsymbol{x}^H \boldsymbol{\Psi}_p^H \boldsymbol{\Psi}_q \boldsymbol{x} & = 
	\boldsymbol{x}^H\boldsymbol{\Psi}_p^H\boldsymbol{y}, \;\;\;\;\; p = 0, \ldots, P-1. 
	\end{aligned}
	\end{equation} 
	Expanding \eqref{eq:likelihood-OTFS} and using the equality \eqref{eq:l-equality}, after some long but relatively simple algebra (not given explicitly for the sake of brevity), we find that the minimization with respect to $\{\tau_p, \nu_p\}$ reduces 
	to maximizing the function
	\begin{align}
	l_2(\boldsymbol{y}|&\boldsymbol{\theta},\boldsymbol{x})=\sum_{p=0}^{P-1}
	\left \{ \underbrace{\frac{|\boldsymbol{x}^H\boldsymbol{\Psi}_p^H\boldsymbol{y}|^2}{\boldsymbol{x}^H\boldsymbol{\Psi}_p^H\boldsymbol{\Psi}_p\boldsymbol{x}}}_{S_p\left(\tau_p,\nu_p\right)}
	- \underbrace{\frac{\left ( \sum_{q\neq p}h'_q \boldsymbol{x}^H \boldsymbol{\Psi}_p^H \boldsymbol{\Psi}_q \boldsymbol{x} \right ) 
	\boldsymbol{y}^H\boldsymbol{\Psi}_p\boldsymbol{x}}{\boldsymbol{x}^H \boldsymbol{\Psi}_p^H \boldsymbol{\Psi}_p\boldsymbol{x}}}_{
	I_p\left(\{h'_q\}_{q\neq p},\boldsymbol{\tau},\boldsymbol{\nu}\right)} \right \} \,,
	\end{align}
	where $S_p$ and $I_p$ denote the useful signal and the interference for path $p$, respectively.
	Clearly, since the channel coefficients $\{h'_p\}$ are not known, it is impossible to directly maximize 
		$l_2\left(\boldsymbol{y}|\boldsymbol{\theta},\boldsymbol{x}\right)$ w.r.t. $\{ \tau_p, \nu_p\}$. Furthermore, 
	even for known coefficients $\{h'_p\}$, the function $l_2\left(\boldsymbol{y}|\boldsymbol{\theta},\boldsymbol{x}\right)$ is not separable in the 
	pairs of parameters $(\tau_p,\nu_p)$ for different values of $p$ because of the dependency of the interference terms $I_p$ on all 
	$(\tau_q,\nu_q)$ for $q \neq p$. Nevertheless, this dependency appears through the ``cross-term'' coefficients of the type
	$\boldsymbol{x}^H \boldsymbol{\Psi}_p^H \boldsymbol{\Psi}_q \boldsymbol{x}$, which tend to be weak for typically 
	sparse multipath channels. Therefore, we resort to an iterative block-wise optimization 
	that alternates the optimization of each pair $(\tau_p,\nu_p)$ by keeping fixed the other parameter pairs and the channel complex coefficients 
	and, after a round of updates for the delay and Doppler parameters, it re-evaluates the estimates of the channel coefficients 
	by solving (\ref{eq:l-equality}).
	The algorithm steps are given in the following, where the iteration index is denoted in brackets by $n^{\mathrm{it}}$:   
	\begin{enumerate}
		\item \textit{Initialization:} let $n^{\mathrm{it}}=0$ and $\hat{h}'_p[0] = 0$ for all $p = 0, \ldots, P-1$. 
		\item For $n^{\mathrm{it}} = 1, 2, 3, \ldots$ repeat: 
		\begin{itemize}
		\item Delay and Doppler shift update:  for each $p = 0, \ldots, P-1$, find the estimates $\hat{\tau}_p[n^{\mathrm{it}}],\hat{\nu}_p[n^{\mathrm{it}}]$ 
		by solving the two-dimensional maximization 
		\begin{eqnarray}
		\left(\hat{\tau}_p[n^{\mathrm{it}}],\hat{\nu}_p[n^{\mathrm{it}}] \right) & = &  
		\arg\max_{\left(\tau_p,\nu_p\right)}   \Big \{ 
		S_p\left(\tau_p,\nu_p\right)  \nonumber \\
		& & - I_p\left(\{\hat{h}'_q[n^{\mathrm{it}}-1]\}_{q\neq p},
		\tau_p,\nu_p, \{ \hat{\tau}_q[n^{\mathrm{it}}-1],\hat{\nu}_q[n^{\mathrm{it}}-1]\}_{q\neq p}\right) \Big \} \nonumber  \\
		& &  \label{diocane}
		\end{eqnarray}
		\item Complex channel coefficients update: solve the linear system (\ref{eq:l-equality}) for the channel matrices 
		$\{\boldsymbol{\Psi}_p\}$ calculated with parameters $\{\hat{\tau}_p[n^{\mathrm{it}}],\hat{\nu}_p[n^{\mathrm{it}}]\}$, and let the solution
		be denoted by $\{\hat{h}'_p[n^{\mathrm{it}}]\}$.
		\end{itemize}
	\end{enumerate}
	The iteration stops when the values of $\{\hat{\tau}_p[n^{\mathrm{it}}],\hat{\nu}_p[n^{\mathrm{it}}]\}$ do not show a significant change with respect to
	the previous iteration, or if a maximum number of iterations is reached. 
	In practice, we find the maximizer in (\ref{diocane}) by searching on a finely 
	discretized grid $\Gamma$ on the delay and Doppler domain. Furthermore, in all our simulations we noticed that the algorithm converges in just a few 
	iterations (2 to 5, at most).  

	\subsection{Performance Bounds}\label{subsec:perf-bounds}
	
	\subsubsection{\acf{CRLB}}

	The derivation of the \ac{CRLB} is based on the channel matrix expression in \eqref{eq:Psi-final}. From the channel model \eqref{eq:h_channel}, by letting
	\begin{equation}
	s_p[k, l]=\left|h'_p\right|e^{j\angle h'_p}\sum_{k'=0}^{N-1}\sum_{l'=0}^{M-1}\boldsymbol{\Psi}^p_{k,k'}\left[l,l'\right]x_{k',l}\,,
	\end{equation}
	and considering separately the amplitude and the phase of the complex channel coefficients $\{h'_p\}$, 
	the $\left(i,j\right)$ element of the $4P\times 4P$ Fisher information matrix is given by 
	\begin{align}\label{eq:fisher-matrix}
	\left[\mathbf{I}(\boldsymbol{\theta},\boldsymbol{x})\right]_{i, j} 
	&= 2\frac{\Pavg}{\sigma_w^2}\mathrm{Re}\left\{\sum_{n,m} \left[\frac{\partial s_p[n,m]}{\partial \theta_i}\right]^*\left[\frac{\partial s_p[n,m]}{\partial \theta_j}\right]\right\},
	\end{align}
	in which $\boldsymbol{\theta}=\left(\left|\boldsymbol{h}'\right|,\angle \boldsymbol{h}',\boldsymbol{\tau},\boldsymbol{\nu}\right)$ is the set of $4P$ channel unknown parameters.
	
	The partial derivatives w.r.t. the magnitude and phase of $h'_p$ are straightforward and are omitted for the sake of space limitation. 	
	The derivatives w.r.t. $\tau_p$ and w.r.t. $\nu_p$ are more cumbersome and, after some algebra, 
	can be obtained in expressions \eqref{eq:Psi-derivative-tau} and \eqref{eq:Psi-derivative-nu}, respectively. 
	\begin{figure*}
		\begin{align}\label{eq:Psi-derivative-tau}
		\frac{\partial\boldsymbol{\Psi}_{k,k'}\left[l,l'\right]}{\partial\tau_p}=\sum_n e^{j2\pi\left(\nu_p NT-k+k'\right)\frac{n}{N}}&\sum_m e^{j2\pi\left(l-l'-\tau_p M\Delta f\right)\frac{m}{M}}
		\left(j2\pi m\Delta f\right)\nonumber\\
		\frac{e^{j2\pi\nu_p\left(\frac{l'}{M\Delta f}\right)}}{NM}
		&\times\begin{cases}
		\begin{array}{ll}
		1 & \mbox{if} \;\; l' \in {\cal L}_{\mathrm{ICI}}\\
		e^{-j2\pi\left(\frac{k'}{N}+\nu_p T\right)} & \mbox{if} \;\; l' \in {\cal L}_{\mathrm{ISI}}
		\end{array}.
		\end{cases}\hspace{-0.2cm}
		\end{align}
		\noindent\rule{\textwidth}{0.4pt}
	\end{figure*}
	\begin{figure*}
		\begin{align}\label{eq:Psi-derivative-nu}
		&\frac{\partial\boldsymbol{\Psi}_{k,k'}\left[l,l'\right]}{\partial\nu_p}=\frac{j2\pi}{NM}\sum_m e^{j2\pi\left(l-l'-\tau_p M\Delta f\right)\frac{m}{M}}e^{j2\pi\nu_p\left(\frac{l'}{M\Delta f}\right)}\nonumber\\
		&\times
		\begin{cases}
		\begin{array}{ll}
		\sum_n e^{j2\pi\left(\nu_p NT-k+k'\right)\frac{n}{N}}\frac{l'}{M\Delta f}+\sum_n nTe^{j2\pi\left(\nu_p NT-k+k'\right)\frac{n}{N}} & \mbox{if} \;\; l' \in {\cal L}_{\mathrm{ISI}}\\
		e^{-j2\pi\left(\frac{k'}{N}+\nu_pT\right)}\left[\sum_n e^{j2\pi\left(\nu_p NT-k+k'\right)\frac{n}{N}}\left(\frac{l'}{M\Delta f}-T\right)+\sum_n nTe^{j2\pi\left(\nu_p NT-k+k'\right)\frac{n}{N}}\right] & \mbox{if} \;\; l' \in {\cal L}_{\mathrm{ISI}}\\
		\end{array}
		\end{cases}
		\end{align}
		\noindent\rule{\textwidth}{0.4pt}
	\end{figure*}
The desired \ac{CRLB} follows by filling the Fisher information matrix in \eqref{eq:fisher-matrix} with the derivatives computed above, and obtaining the diagonal elements of the inverse Fisher information matrix. In particular, we are interested in the \ac{CRLB} for the parameters $\tau_0$ and $\nu_0$, 
related to the target range and velocity.  
	
	\subsubsection{Waterfall Analysis for $P=1$}
	
	In the case of $P = 1$ our iterative scheme reduces to the exact joint ML estimation of $h'_0, \tau_0, \nu_0$ (in other words, the iterative scheme converges 
	after the first iteration to the exact maximum of the log-likelihood function, up to the discretization error in the 2-dimensional search domain $\Gamma$).
	It is well-known that \ac{ML} estimators typically exhibit a threshold effect, i.e., a rapid deterioration of the estimation \ac{MSE} when the \ac{SNR} is below some threshold (that generally depends on the problem and on the sample size). In contrast, for \ac{SNR} larger than such threshold the ML
	estimator yields MSE typically very close to the \ac{CRLB}.
This waterfall behavior is caused by ``outliers'' in the search of the maximum in \ac{ML} estimator: namely, when the observation is too noisy, 
the maxima of the likelihood functions tend to be randomly placed anywhere on the search grid. 
In the following we provide an analysis of the waterfall transition for the case $P = 1$, i.e., 
the region around the threshold \ac{SNR} value where the rapid deterioration  occurs. Our simulations show that the waterfall prediction provided by our analysis for $P = 1$ is also very accurate for the multipath case $P > 1$. This corroborates the evidence that our proposed approximated ML algorithm is effectively 
very good, and performs very close to the true ML.\footnote{Obviously, the CRLB for $P = 1$ yields also a lower bound for the case $P > 1$, in the case where
we simply add more multipath components on top of the \ac{LoS} component without changing the statistics of the \ac{LoS} component, since the presence of 
more unknown parameters cannot help the estimation of the target parameters $\tau_0,\nu_0$.} 

Following the reasoning of \cite{athley2003space,athley2005threshold}, we treat as ``outlier'' the event that a maximum of the log-likelihood function 
is randomly placed on the grid $\Gamma$, rather than in the cluster gird points around the true value $(\bar{\tau}_0,\bar{\nu}_0)$. 
Let $\alpha \in \left\{\tau,\nu\right\}$ be the unknown parameter to be estimated. 
By the law of total probability over the discretized grid $\Gamma$ (where we calculate the \ac{ML} estimator), 
we can approximate the estimation MSE as
	\begin{align}\label{eq:MSE-waterfall}
	\mathrm{MSE}=E\left[\left(\hat{\alpha}-\bar{\alpha}\right)^2\right]&=\sum_{i\in\Gamma}\Pr\left(\epsilon\left(i\right)\right)\left(\hat{\alpha}_i-\bar{\alpha}\right)^2\leq\sum_{i\in\Gamma}\Pr\left(\tilde{\epsilon}\left(i\right)\right)\left(\hat{\alpha}_i-\bar{\alpha}\right)^2\,.
	\end{align}
	where $\bar{\alpha}$ is the true value of the parameter, $\hat{\alpha}$ is the estimated parameter, and $\Pr\left(\epsilon\left(i\right)\right)$ denotes the probability of error of choosing $\alpha_i$ rather than $\bar{\alpha}$.	While evaluating $\Pr\left(\epsilon\left(i\right)\right)$ may be extremely difficult, we obtain an upper bound by considering pairwise error probabilities, i.e., replacing 
	$\Pr\left(\epsilon\left(i\right)\right)$ with the probability that the detector chooses $\alpha_i$ rather than $\bar{\alpha}$ when these are the only two alternatives.	Since the pairwise error event $\tilde{\epsilon}\left(i\right)$ contains the 
	true error event $\epsilon(i)$, it follows that the inequality of \eqref{eq:MSE-waterfall} provides an upper bound. 
	
	Proceeding further, we define the pairwise error probability as
	\begin{equation}
	\Pr\left(\tilde{\epsilon}\left(i\right)\right)\triangleq \Pr\Big\{l\left(\boldsymbol{y}|\boldsymbol{\theta}_i,\boldsymbol{x}\right)<l\left(\boldsymbol{y}|\bar{\boldsymbol{\theta}},\boldsymbol{x}\right)\Big\}\,,
	\end{equation}
	where $l\left(\boldsymbol{y}|\boldsymbol{\theta}_i,\boldsymbol{x}\right)$ and $l\left(\boldsymbol{y}|\bar{\boldsymbol{\theta}},\boldsymbol{x}\right)$ are the value obtained from the evaluation of the likelihood function at grid points $i$ with parameters $\boldsymbol{\theta}_i$ and using the true parameters $\bar{\boldsymbol{\theta}}$, respectively. 
	At this point, the problem is reduced to the computation of the pairwise error probabilities $\Pr\left(\epsilon\left(i\right)\right)$, for $i\in\Gamma$, which can be derived as follows. We notice that
	\begin{align}
	\Pr\left(\tilde{\epsilon}\left(i\right)\right)\approx\Pr\left\{| \boldsymbol{x} ^H \boldsymbol{\Psi}^H_i \boldsymbol{y}|^2>| \boldsymbol{x} ^H \bar{\boldsymbol{\Psi}}^H \boldsymbol{y}|^2\right\}\,,
	\end{align}
	where $\bar{\boldsymbol{\Psi}}$ is the channel matrix calculated with the true parameters and we exploit $\boldsymbol{\Psi}^H_i\boldsymbol{\Psi}_i\simeq\boldsymbol{I}$, $\forall i$, for $P=1$. Defining the jointly conditionally Gaussian random variables
	\begin{eqnarray}
	z_i & \triangleq & \boldsymbol{x}^H\boldsymbol{\Psi}^H_i\boldsymbol{y}=\boldsymbol{x}^H\boldsymbol{\Psi}^H_i\bar{\boldsymbol{\Psi}}\boldsymbol{x}+\boldsymbol{x}^H\boldsymbol{\Psi}^H_i\boldsymbol{w} \\
	\bar{z} & \triangleq & \boldsymbol{x}^H\bar{\boldsymbol{\Psi}}^H\boldsymbol{y}\approx\boldsymbol{x}^H \boldsymbol{x}+\boldsymbol{x}^H\bar{\boldsymbol{\Psi}}^H\boldsymbol{w}\,,
	\end{eqnarray}
	with first and second order moments
	\begin{equation}\label{eq:first-second-moments}
	\begin{cases}
	\begin{array}{ll}
	\mathrm{E}\left[\bar{z}\right]=\left\|\boldsymbol{x}\right\|^2,  & \mathrm{Var}\left[\bar{z}\right]=\sigma_w^2\left\|\boldsymbol{x}\right\|^2\\ \mathrm{E}\left[z_i\right]=\boldsymbol{x}^H\boldsymbol{\Psi}_i^H\bar{\boldsymbol{\Psi}}\boldsymbol{x},  & \mathrm{Var}\left[z_i\right]=\sigma_w^2\left\|\boldsymbol{x}\right\|^2\\
	\end{array}\,,
	\end{cases}
	\end{equation}
	and
	\begin{equation}
	\mathrm{Cov}\left[\bar{z},z_i\right]=\sigma_w^2 \boldsymbol{x}^H\bar{\boldsymbol{\Psi}}^H\boldsymbol{\Psi}_i\boldsymbol{x}\, ,
	\end{equation}
	a good approximation for $\Pr\left(\tilde{\epsilon}\left(i\right)\right)$ is given by \cite{athley2003space}
	\begin{equation}\label{eq:P_i-approx}
	\Pr\left(\tilde{\epsilon}\left(i\right)\right)\approx\frac{1}{2}\exp\left\{-\frac{\left\|\boldsymbol{x}\right\|^4}{2\sigma_w^2NM} \right\}\mathrm{I}_0\left(\frac{\left|\boldsymbol{x}^H\boldsymbol{\Psi}^H_i\bar{\boldsymbol{\Psi}}\boldsymbol{x}\right|\cdot\left\|\boldsymbol{x}\right\|^2}{2\sigma_w^2NM}\right)\,,
	\end{equation}
	where $\mathrm{I}_0$ is the modified Bessel function of the first kind of order $0$. 
	The \ac{MSE} estimation can be thus computed by substituting \eqref{eq:P_i-approx} into \eqref{eq:MSE-waterfall}, using the above definitions.

	The resulting (approximated) upper bound tends to be loose at low \ac{SNR}, but becomes more accurate moving towards the \ac{ML} waterfall region.
	Furthermore, the \ac{MSE} strictly depends on the grid resolution and may or may not reach the CRLB for increasing SNR 
	depending on the systematic error incurred by the grid discretization. As a matter of fact, in our numerical results we took care of using a 
	search grid fine enough such that the discretization systematic error is not visible in the explored range of SNR. 
	
	In order to cope with the fact that the (approximated) MSE upper bound obtained in \eqref{eq:MSE-waterfall} becomes very loose for low SNR, we 
	use also the trivial upper bound 
	\begin{equation}\label{eq:random-estimation}
		\mathrm{MSE} \leq \frac{1}{\left|\Gamma\right|}\sum_{i\in\Gamma}(\hat{\alpha}_i-\bar{\alpha})^2\,,
	\end{equation} 
	that corresponds to choosing at random grid points, by disregarding completely the received signal. 
	
	Then, putting together \eqref{eq:MSE-waterfall} (with the pairwise error probability approximation in \eqref{eq:P_i-approx} 
	and the above ``random estimation'' bound in (\ref{eq:random-estimation}), we finally obtain the approximated MSE upper bound
	\begin{equation}
		\mathrm{MSE} \lesssim \min\left\{\sum_{i\in\Gamma}\Pr\left(\tilde{\epsilon}\left(i\right)\right)\left(\hat{\alpha}_i-\bar{\alpha}\right)^2, \frac{1}{\left|\Gamma\right|}\sum_{i\in\Gamma}(\hat{\alpha}_i-\bar{\alpha})^2 \right\}\,.
	\end{equation}
	Our simulations show that this approximated upper bound is very accurate and it is able to predict very well the waterfall behavior of the 
	ML estimator (as well as the proposed approximated ML estimator in the case $P > 1$).

	\section{OTFS Soft-Output Data Detection}\label{sec:Detection-Algorithm}
	
	We will now focus on the \ac{OTFS} data detection at the receiver side. 
	By considering the communication channel model in \eqref{eq:h_channel}, in line with most of the current literature on \ac{OTFS} detection (e.g., \cite{raviteja2018interference,hadani2017otfs}), we assume perfect \ac{CSI} at the receiver. 
	An efficient pilot-aided \ac{CSI} acquisition is a problem of independent interest that we postpone to future work (see also \cite{raviteja2019OTFSchannelEst,shen2019channel,zhang2018channelEstOTFScs}). 
	
	In this paper, we consider separate detection and decoding, where the receiver consists of the concatenation of a
	soft-output symbol detector, producing soft-estimates of the coded symbols $\boldsymbol{x}$, and a decoder that takes such 
	estimates as the output of a virtual channel that incorporates also the detector. Motivated by practical complexity considerations, 
	no ``turbo equalization'' involving a feedback loop from the (soft-output) decoder to the detector and iterations between 
	detector and decoder is considered. It follows that the relevant performance measure is the already mentioned
pragmatic capacity, i.e., the mutual information between the input constellation symbols, used with uniform probability, 
and the corresponding detector soft-output \cite{kavcic2003binary,soriaga2007determining}.	
	
We propose an efficient low-complexity \ac{MP}-based soft-output detector, obtained by 
constructing a \ac{FG} for the joint posterior probability of $\boldsymbol{x}$ given $\boldsymbol{y}$ in \eqref{eq:y} and applying the standard 
\ac{SPA} computation rules to compute its marginals. The \ac{FG} is constructed according to the general approach of \cite{colavolpe2011SISOdetection} (applicable to 
any linear-Gaussian model such as \eqref{eq:y}), for which the graph girth is guaranteed to be at least 6 and 
the degree of the function nodes is at most 2. 
This allows the application of the exact \ac{SPA} computation at the nodes and a high degree of parallelization, such that the resulting \ac{MP}-based 
detector is very computationally efficient. 
Furthermore the absence of cycles of length less than 6 yields good convergence properties of the 
\ac{SPA} iterations. 

We compare the proposed scheme with other three soft-output detectors proposed in the literature, namely:
i) a different \ac{MP}-based detector recently proposed in \cite{raviteja2018interference}, based on an alternative way to construct the  
\ac{FG} (see Section \ref{algo-Psi});
ii) a standard linear MMSE block equalizer, which offers very good performance at the impractical cost of a very large-dimensional matrix inversion;
iii) a low-complexity MMSE block equalizer recently proposed in \cite{cheng2019low} that uses drastic simplifying assumptions (in particular, 
the bi-orthogonality of the OTFS $\gTX$ and $\gRX$ pulses and the assumption that 
the delay and Doppler shifts as integer multiples of the receiver sampling grid) such that the resulting 
{\em nominal} channel matrices  $\boldsymbol{\Psi}_p$ are block-circulant with circulant blocks and the matrix inversion in 
linear MMSE estimation can be efficiently implemented. As we shall see, since these simplifying assumptions are not verified in practice
(shifts are not on a quantized grid, and bi-orthogonal pulses with unit time-frequency product cannot exist \cite{matz2013time}), 
the performance of this low-complexity linear detector is quite poor, when applied to a realistic channel and pulse scenario.



\subsection{Proposed MP-based detector (``Matrix $\boldsymbol{G}$ algorithm'' --- $\mathrm{MP}_{\boldsymbol{G}}$)}
\label{algo-G}

From \eqref{eq:y}, when $\boldsymbol{\Psi}$ is known (perfect CSI) and $\boldsymbol{w}$ is complex Gaussian i.i.d., 
the conditional \ac{pdf} of the received samples $\boldsymbol{y}$ given the modulation symbols $\boldsymbol{x}$ is given by  
	\begin{equation}
		p\left(\boldsymbol{y}|\boldsymbol{x}\right)=\frac{1}{\left(2\pi\sigma_w^2\right)^{-NM}}\exp\left(-\frac{\left\|\boldsymbol{y}-\boldsymbol{\Psi x}\right\|^2}{2\sigma_w^2}\right)\propto\exp\left(-\frac{\left\|\boldsymbol{y}-\boldsymbol{\Psi x}\right\|^2}{2\sigma_w^2}\right)\,,  \label{ziominchia}
	\end{equation}
	where the proportionality symbol indicates an irrelevant constant factor independent of symbols $\boldsymbol{x}$. We follow the \ac{FG} construction approach 
of \cite{colavolpe2011SISOdetection}, and expand the $\ell_2$-norm inside the exponential as
	\begin{equation}
		\left\|\boldsymbol{y}-\boldsymbol{\Psi x}\right\|^2=\boldsymbol{y}^H\boldsymbol{y}-2\mathrm{Re}\left\{\boldsymbol{x}^H\boldsymbol{\Psi}^H\boldsymbol{y}\right\}+\boldsymbol{x}^H\boldsymbol{\Psi}^H\boldsymbol{\Psi}\boldsymbol{x}.
	\end{equation}
	Defining $\boldsymbol{z}\triangleq\boldsymbol{\Psi}^H\boldsymbol{y}$ and $\boldsymbol{G}\triangleq\boldsymbol{\Psi}^H\boldsymbol{\Psi}$, 
	the conditional \ac{pdf} can be written as
	\begin{equation}
		p\left(\boldsymbol{y}|\boldsymbol{x}\right)\propto\exp\left(\frac{2\mathrm{Re}\left\{\boldsymbol{x}^H\boldsymbol{z}\right\}-\boldsymbol{x}^H\boldsymbol{G}\boldsymbol{z}}{2\sigma_w^2}\right)\,.
	\end{equation}
	Note that the sequence $\boldsymbol{z}$ is a sufficient statistic for symbols detection.  
	By expressing the matrix operations explicitly in terms of the components, we define the functions
	\begin{equation}\label{eq:F-function}
		F_i\left(x_i\right)=\exp\left[\frac{1}{\sigma_w^2}\mathrm{Re}\left\{z_ix_i^*-\frac{G_{i,i}}{2}\left|x_i\right|^2\right\}\right]\,,
	\end{equation}
	\begin{equation}\label{eq:I-function}
		I_{i,j}\left(x_i,x_j\right)=\exp\left[-\frac{1}{\sigma_w^2}\mathrm{Re}\left\{G_{i,j}x_jx_i^*\right\}\right]\,,
	\end{equation}
	and we use Bayes rule in order to express the a-posteriori probability of $\boldsymbol{x}$ given $\boldsymbol{y}$ in the factored form, i.e., 
	\begin{equation}\label{eq:G-factorization}
		P\left(\boldsymbol{x}|\boldsymbol{y}\right)\propto P\left(\boldsymbol{x}\right)p\left(\boldsymbol{y}|\boldsymbol{x}\right)\propto\prod_{i=1}^{NM}\left[P\left(x_i\right)F_i\left(x_i\right)\prod_{j<i}I_{i,j}\left(x_i,x_j\right)\right]\,,
	\end{equation}
where we used the fact that the modulation symbols $x_i$ take on values in some signal constellation $\mathcal{C}$ 
and are treated by the detector as i.i.d. with given (typically uniform) a priori probability mass function  $\{P(x) : x \in \mathcal{C}\}$.

\begin{figure}
		\centering
		\includegraphics[scale=0.8]{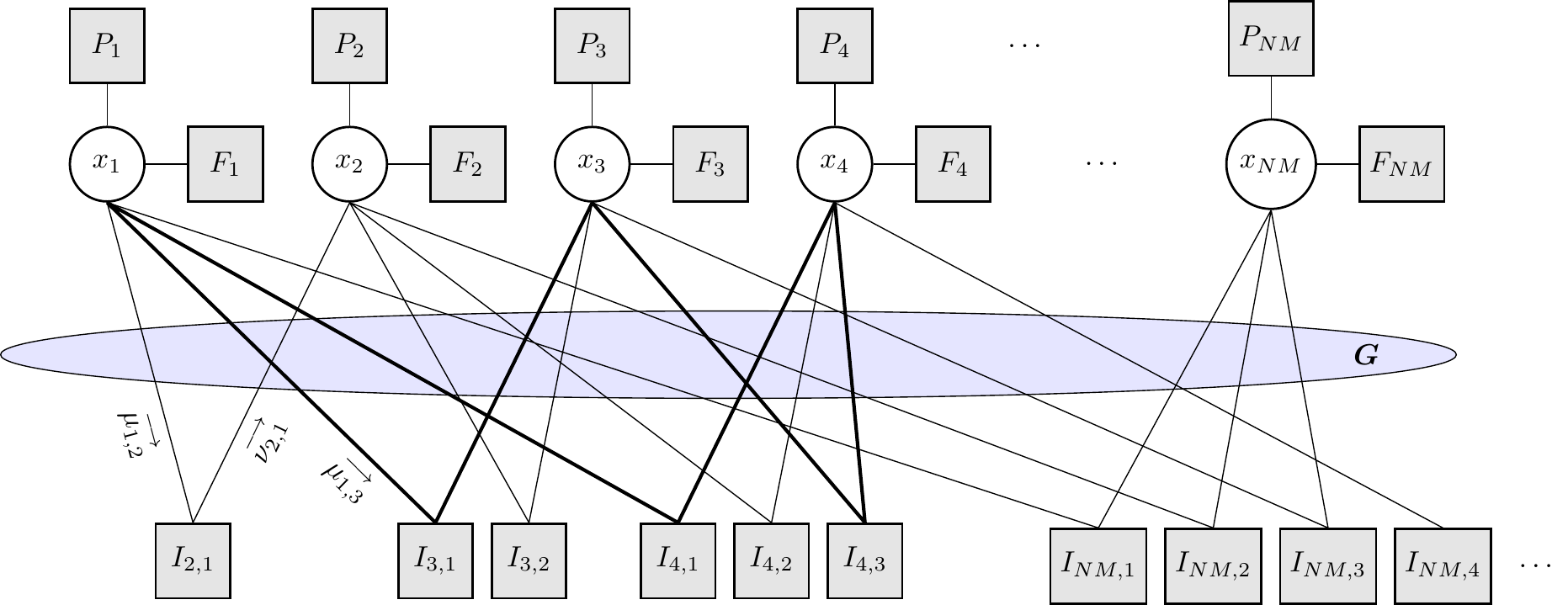}
		\caption{Structure of the FG for the $\mathrm{MP}_{\boldsymbol{G}}$ algorithm.}
		\label{fig:FG_G}
	\end{figure}

	In the proposed approach, the \ac{FG} corresponds to the factorization in \eqref{eq:G-factorization} (see the example 
	shown in Fig. \ref{fig:FG_G}). At this point, the resulting \ac{MP}-based soft-output detector follows immediately 
	by applying the standard \ac{SPA} computation rules. The detailed derivation of the message computation at the function and variable nodes of the 
	\ac{FG}  is given in  \cite{colavolpe2011SISOdetection} and applies directly to our setting. 
	Here, for the sake of completeness, we just summarize the resulting algorithm. 
Define $V_i\left(x_i\right)$ as the product of all messages incoming to the variable node $x_i$, namely
	\begin{equation}
		V_i\left(x_i\right)=P_i\left(x_i\right)F_i\left(x_i\right)\prod_{j\neq i}\nu_{i,j}\left(x_j\right)\,,
	\end{equation}
	which is proportional to the (estimated) a-posteriori probability $P\left(x_i|\boldsymbol{y}\right)$ and thus provides the 
	soft-output of the this detector.	Then, the application of the \ac{SPA} leads to the following rules for message exchange and update: 
	\begin{enumerate} 
	\item Computation at the variable nodes: each node $x_i$ sends to each adjacent function node $I_{i,j}$ the message 
	\begin{equation}
		\mu_{i,j}\left(x_i\right)=V_i\left(x_i\right)/\nu_{i,j}\left(x_i\right).
	\end{equation}	
	\item Computation at the function nodes: each function node $I_{i,j}$ sends to each adjacent variable node $x_i$ the message
	\begin{equation}
		\nu_{i,j}\left(x_i\right)=\sum_{x_j \in \mathcal{C}}I_{i,j}\left(x_i,x_j\right)\mu_{i,j}\left(x_j\right)\,.
	\end{equation}
	\end{enumerate}
	Notice that all the variable nodes and the function nodes can be activated in alternative rounds and, in each round, all the nodes of the same 
	type can be activated in parallel, e.g., adopting the same flooding schedule used for \ac{LDPC} decoding \cite{Richardson2001capacity}. Moreover, messages can be implemented in the logarithmic domain \cite{Kschischang2001SPA}. 

	With the help of Fig. \ref{fig:FG_G}, we can illustrate some important features of the proposed approach. 
	\begin{itemize}
		\item[i)] FG of girth 6: the \ac{FG} constructed as above is guaranteed to have girth (minimum length of cycles) equal to 6 
		(highlighted in bold in Fig. \ref{fig:FG_G}). It is well-known that the SPA yields exact posterior marginalization for cycle-free \acp{FG}, and the rationale behind the use of the SPA paradigm on loopy graphs
		is that if the FG has large girth, the local neighborhood of each node is ``tree-like'' \cite{Kschischang2001SPA}. In particular, 
		cycles of length 4 should be avoided. Hence, the proposed construction following the general method of  \cite{colavolpe2011SISOdetection} yields
		indeed an FG better suited to the application of the iterative SPA. 
		\item[ii)] Computational complexity: notice that the computation in \eqref{eq:I-function} involves the summation with respect to only a single discrete variable over the constellation $\mathcal{C}$. Therefore, this computation has always linear complexity in 
		the constellation size, irrespectively on the sparsity of the channel matrix $\boldsymbol{\Psi}$. 
		This allows the use of the exact \ac{SPA} at fixed complexity per node, unlike the approach in \cite{raviteja2018interference}  
		(see comments in Section \ref{algo-Psi}). 
		\item[iii)] High degree of parallelization: the number of nodes of type $I_{i,j}$ depends on the number of non-zero elements in the 
		rows of the upper triangular part of the matrix $\boldsymbol{G}$ (the existence of edges is evidenced by the shadowed elliptic area in Fig. \ref{fig:FG_G}). Nevertheless, as said before, these degree-2 nodes $I_{i,j}$ can be all activated in parallel. Hence, for a sufficiently large degree of parallelization, 
		the computational {\em time complexity} is independent of the sparsity of the multipath channel. Notice that in modern LDPC decoding is not 
		unlikely to find implementations with degree of parallelization of the order of 1000, which is much larger than what needed in our detector. 
		Hence, we claim that the proposed detector is very attractive from a practical implementation viewpoint. 
\end{itemize}


	\subsection{MP-based algorithm of \cite{raviteja2018interference} (``Matrix $\boldsymbol{\Psi}$ algorithm'' --- $\mathrm{MP}_{\boldsymbol{\Psi}}$)}
	\label{algo-Psi}
	
	\begin{figure}
		\centering
		\includegraphics[scale=1]{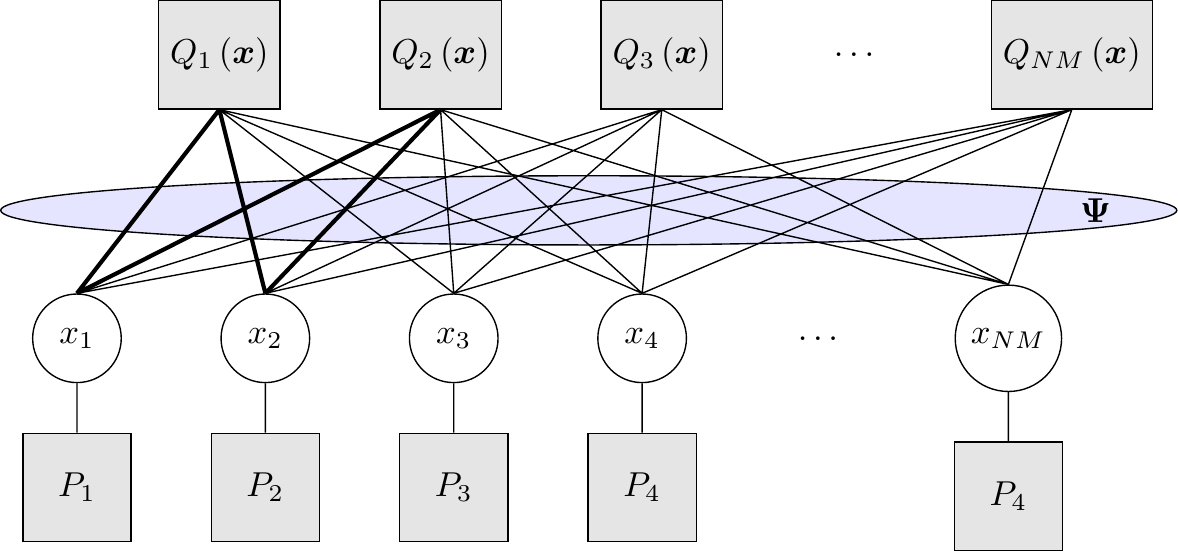}
		\caption{Structure of the FG for the $\mathrm{MP}_{\boldsymbol{\Psi}}$ algorithm.}
		\label{fig:FG_Psi}
	\end{figure}

	The \ac{MP}-based algorithm of \cite{raviteja2018interference} builds its \ac{FG} from the more ``direct'' factorization of the a-posteriori probability 
	\begin{equation}
		P\left(\boldsymbol{x}|\boldsymbol{y}\right)\propto P\left(\boldsymbol{x}\right) p\left(\boldsymbol{y}|\boldsymbol{x}\right)
		=\prod_{i=1}^{N\times M} p\left(y_i|\boldsymbol{x}\right) \prod_{i=1}^{N\times M} P\left(x_i\right).
	\end{equation}
	Defining the function nodes
	\begin{equation}
	Q_i\left(\boldsymbol{x}\right) \triangleq p\left(y_i|\boldsymbol{x}\right)=\exp\left(\frac{\left\|y_i-\boldsymbol{\Psi}_i\boldsymbol{x}\right\|^2}{2\sigma_w^2}\right)\,,
	\end{equation}
	where $\boldsymbol{\Psi}_i$ is the $i$-th row of $\boldsymbol{\Psi}$, the resulting \ac{FG} is shown in  Fig. \ref{fig:FG_Psi}. 
	Since the elements of the same $i$-th column $\boldsymbol{\Psi}$ is multiplied by the same symbol $x_i$, 
	this \ac{FG} has necessarily cycles of length 4 (highlighted in bold in Fig. \ref{fig:FG_Psi}). 
	The edges evidenced by the shadowed elliptic area in the figure correspond 
	to the non-zero elements of the matrix $\boldsymbol{\Psi}$. 
	In particular, the degree $d_i$ of the function node $Q_i(\boldsymbol{x})$ is equal to the number of 
	non-zero elements in the the $i$-th row $\boldsymbol{\Psi}_i$. The exact SPA computation at such function nodes requires summing over
	$d_i - 1$ discrete variables taking values in $\mathcal{C}$. Therefore, it has complexity $|\mathcal{C}|^{d_i-1}$ that may be 
	prohibitively large for large constellations and, above all, it depends on the sparsity of the channel. 
	Therefore, the exact application of the  SPA computation rules to the FG obtained directly from the matrix $\boldsymbol{\Psi}$ 
	is highly impractical. For this reason, the authors of \cite{raviteja2018interference} propose to use a Gaussian approximation of the interfering symbols 
	in the computation at the nodes $Q_i\left(\boldsymbol{x}\right)$, which effectively boils down to a soft interference cancellation approach, 
	as already widely used in turbo equalization and pioneered in the context of multiuser detection in \cite{wang1999interference,boutros2002multiuserDecoding}.
For the sake of space limitation, we omit the details of the resulting \ac{MP} algorithm, which can be found in \cite{raviteja2018interference}.
	
It should also be mentioned that, for the sake of simplicity and in order to increase the sparsity of the \ac{FG} in Fig. \ref{fig:FG_Psi}, 
the detector proposed in \cite{raviteja2018interference} constructs the {\em nominal} matrix $\boldsymbol{\Psi}$ by rounding the delay 
shifts  to integers on receiver sampling grid. Under this condition, the channel matrix is very sparse since many coefficients 	
corresponding to sampling at non-integer delay shifts are identically to zero, and the number of connections for each node is 
reduced, while preserving the 4-cycle problem said before since this is unavoidable with this approach. 
Nevertheless, since the assumption is generally not satisfied by real-world channels, such approximation of the channel matrix results in neglecting 
a significant component of the \ac{ISI}. We shall verify that when such integer delay shift rounding is applied to the construction 
of the nominal matrix $\boldsymbol{\Psi}$ used by the detector, but the actual delay shifts have a fractional component (as it is always the case in practice), this {\em mismatch} yields a significant performance degradation.
This shows that neglecting the fractional part of delays and Doppler shifts, as routinely done in the literature of OTFS, may be 
indeed quite misleading.
	

\subsection{Linear block-wise MMSE equalization}
\label{LMMSEeq}

As a further term of comparison we consider also the standard linear MMSE block equalizer, applied to the channel model \eqref{eq:y}.
In this case, the soft-output is simply the linear MMSE estimate of symbols $\boldsymbol{x}$ from the observation 
$\boldsymbol{y}$, given by 
	\begin{equation}  \label{lmmse-eq}
		\hat{\boldsymbol{x}}_{\mathrm{LMMSE}}=\boldsymbol{\Psi}^H\left(\boldsymbol{\Psi}\boldsymbol{\Psi}^H+\sigma_w^2\boldsymbol{I}\right)^{-1}\boldsymbol{y}\,.
	\end{equation}
	The complexity of this approach is proportional to $\mathcal{O}\left(\left(NM\right)^3\right)$, so it becomes quickly unfeasible for typical values of $N$ and $M$
	(e.g., in our simulations we have considered $N=50$ and $M=64$).  
	A low-complexity (mismatched) \ac{LMMSE} approach was 
	recently proposed in \cite{cheng2019low}. This relies on the cyclic properties of channel matrix $\boldsymbol{\Psi}$ under 
	perfect bi-orthogonality of the modulation pulses $\gTX$ and $\gRX$ and integer-grid valued delays and Doppler shifts.
	As a result, in the presence of practical rectangular pulses non-integer delays and Doppler shifts, as realistically considered in our work,
	the performance of this approach visibly degrades.

	\section{Simulation Results}\label{sec:sim-results}
	
		The radar (backscattered) and forward communication channels are both defined by \eqref{eq:h_channel} 
	with different values of \ac{PL}, Doppler shift, and delay. We define the radar and communication SNRs as  \cite[Chapter 2]{richards2014fundamentals}
\begin{equation}
\mathrm{SNR_{rad}}=\frac{\lambda^2\sigma_{\mathrm{rcs}}G^2}{\left(4\pi\right)^3r^4}\frac{\Pavg}{\sigma_w^2}\,, \;\;\; 
\mathrm{SNR_{com}}=\frac{\lambda^2G^2}{\left(4\pi\right)^2r^2}\frac{\Pavg}{\sigma_w^2}\,,
\end{equation}
respectively,  where $\lambda=c/f_c$ is the wavelength, $\sigma_{\mathrm{rcs}}$ is the radar cross-section 
	in $\mathrm{m}^2$, $G$ is the antenna gain, and $r$ is the distance between transmitter and receiver. 
	
	In the case of multipath, we fix $\mathrm{SNR_{com}}$ to be the SNR of the \ac{LoS} component, and we add 
	multipath components with progressively lower SNRs, such that the sum SNR of the channel increases with the number of paths $P$. 
	This corresponds to the physically meaningful case that a richer propagation 
	environment conveys more signal power. 	Table \ref{table:parameters} summarizes the relevant simulation parameters 
	inspired by the automotive communication standard IEEE 802.11p \cite{nguyen2017delay}, where $r$ and $v$ denote the target range and velocity.

	\begin{table}
		\centering
		\renewcommand*{\arraystretch}{1.5}
		\caption{Simulation parameters}
		\begin{tabular}{|c|c|}
			\hline
			$f_c=5.89$ GHz & $M=64$ \\ \hline
			$B=10$ MHz & $N=50$ \\ \hline
			$\Delta f=B/M=156.25$ kHz & $T=1/\Delta f=6.4\,\mu s$ \\ \hline\hline
			$\sigma_{\mathrm{rcs}}=1$ m\textsuperscript{2} & $G=100$ \\ \hline
			$r=20$ m & $v=80$ km/h \\ \hline
		\end{tabular}
		\label{table:parameters}
	\end{table}

\subsection{Comparison with OFDM and FMCW}
\newcommand{\Tg}{T_{\rm GI}}
\newcommand{\eqdef}{\stackrel{\Delta}{=}}

We briefly review \ac{OFDM} and the widely used radar waveform known as \ac{FMCW} \cite[Chapter 4.6]{richards2014fundamentals}.
For both \ac{OFDM} and \ac{FMCW} we consider a symbol length of $T_0= \Tg +T$ including a guard interval denoted by $\Tg$, longer than 
the maximum path delay $\tau_{\max}$. In OFDM, the \ac{CP} length is given by $\Tg=C\frac{T}{M}$, with $C=\lceil \frac{\tau_{\max}}{T/M}\rceil$. 
We send $MN$ modulation symbols $\{x_{n, m}\}$ in the time-frequency domain, satisfying the average power constraint \eqref{eq:avg-power}. 
The OFDM receiver samples at rate $T/M$ and removes the CP before detection in order to eliminate the inter-symbol 
interference.  
For the case of a single path channel ($P=1$), the ML estimator and the related \ac{CRLB} can be found in
our related work \cite{gaudio2019performance} and are omitted here for the sake of space limitation. 

In \ac{FMCW}, the radar transmitter sends a sequence of identical ``chirp'' pulses of duration $T$, each followed by a guard interval of $\Tg$ to avoid inter-pulse interference. By letting $\phi(t) = ( f_c +  \frac{B}{2T}t )t$ denotes the phase at time $t$, 
the transmit signal is given by  
\begin{align}\label{eq:FMCW-input}
s\left(t\right) = \sum_{i=0}^{N-1}e^{j 2\pi \phi(t-iT_0)} \rect\left(\frac{t-iT_0}{T}\right)\,,
\end{align} 
where we consider $N$ pulses to make a fair comparison with OFDM. Plugging \eqref{eq:FMCW-input} into \eqref{eq:OTFS-RxSignal} , we obtain the received signal $r(t)$ by neglecting the noise. 
After some algebra, it is easy to show that the product of the received and the transmit signals gives
\begin{align}
y(t) &= r(t)s^*(t)
= \sum_{p=0}^{P-1} h_p  e^{j 2\pi \nu_p t} e^{-j 2\pi f_c \tau_p}\sum_{i=0}^{N-1}   e^{-j 2\pi f_{b,p} (t-i T_0 -\tau_p/2)}\,,
\end{align}
where we let 
$f_{b,p} = \frac{B}{T} \tau_p $ denotes the so-called {\em beating frequency} of path $p$. 
The receiver samples $y(t)$ every $T/M$ for each pulse, i.e. for $t= i T_0 + l \frac{T}{M}$ where $i$ denotes the pulse index and $l$ denotes the sample index. 
By letting $L=M+C$ the number of samples par pulse, the sampled received signal can be rewritten as
\begin{align}\label{eq:received-samples}
y[i, l] &\eqdef y\left(i T_0 + \frac{l}{f_s} \right)
= \sum_p h_p   e^{j 2\pi (f_{b,p}+\nu_p)  \frac{l}{f_s} }e^{j 2\pi \nu_p i T_0}\,, 
\end{align}
for $i = 0, \ldots, N-1$ and $l = 0, \ldots, L-1$, where $h_p$ absorbs a constant phase term independent of the indices $i, l$.   

The estimation of the $2P$ unknown parameters $\{\tau_p,\nu_p\}$ is thus obtained by selecting the peak of the range-Doppler map found by applying a two-dimensional \ac{DFT} to the noisy samples in \eqref{eq:received-samples} as proposed in \cite{braun2014ofdm,patole2017automotive}. 
The range and velocity of the target are obtained by the estimates of $\{\tau_0, \nu_0\}$.

\subsection{Joint Radar and Communication Performance}

\begin{figure}
		\centering
		\includegraphics[scale=0.8]{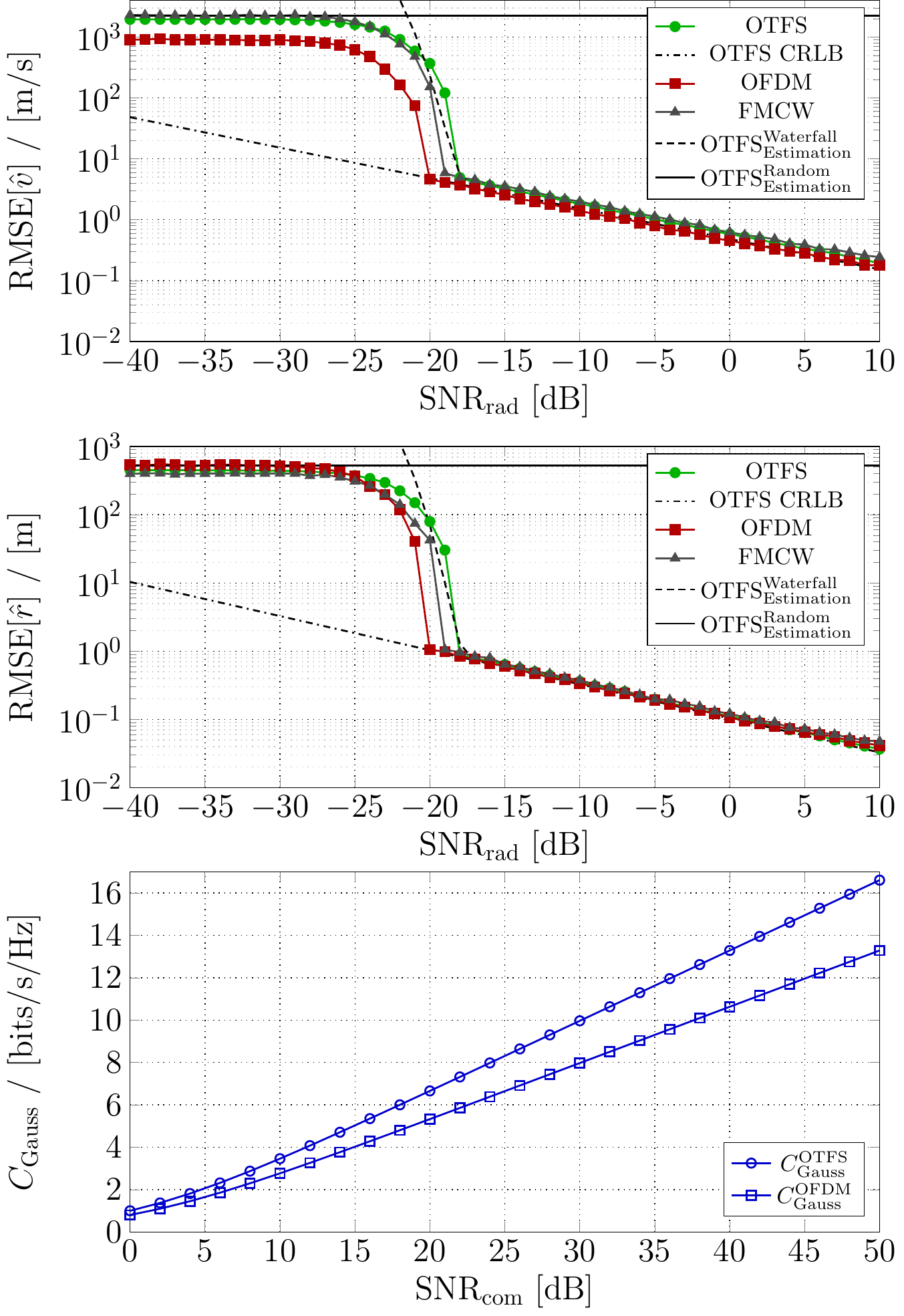}
		\caption{From top to bottom: the RMSE of the target velocity estimation $\hat{v}$ vs $\mathrm{SNR_{rad}}$, the RMSE of the target range estimation $\hat{r}$ vs $\mathrm{SNR_{rad}}$, and the Gaussian capacity $\mathcal{C}_\mathrm{Gauss}$ vs. $\mathrm{SNR_{com}}$, 
		for the curves indicated in the different legends.
		}
		\label{fig:RMSE-vel-range-rate}
	\end{figure}

The first two subfigures of Fig.~\ref{fig:RMSE-vel-range-rate} show the 	
velocity and range estimation root \ac{MSE} (RMSE) versus $\mathrm{SNR_{rad}}$ for a pure \ac{LoS} channel ($P=1$) and for 
\ac{OTFS}, \ac{OFDM}, and \ac{FMCW}. We notice that both digital modulation formats provide 
as accurate radar performance as \ac{FMCW}, while transmitting at their full information rate. 
It is remarkable to note that the combination of the waterfall analysis and the \ac{CRLB} introduced in 
Section \ref{subsec:perf-bounds} are able to accurately predict the actual \ac{ML} estimation performance. 

In addition, the third subfigure of Fig. \ref{fig:RMSE-vel-range-rate} shows the achievable rate with Gaussian independent and identically 
distributed inputs symbols $C_\mathrm{Gauss}$, which provides an achievable rate in the case of {\em joint detection and decoding} 
(with unconstrained complexity), as a function of $\mathrm{SNR_{com}}$. 
Given the fact that we can model the input-output \ac{OTFS} block channel as a \ac{MIMO} channel, 
the mutual information with Gaussian inputs and perfect \ac{CSI} at the receiver is given by \cite{telatar1999capacity}
	\begin{equation}\label{eq:57}
		C_\mathrm{Gauss}^\mathrm{OTFS} = \frac{NT}{NT + T_{\rm GI}} \left ( \frac{1}{NM} \log_2\det\left(\boldsymbol{I}+\mathrm{SNR_{com}}\boldsymbol{\Psi}\boldsymbol{\Psi}^H\right) \right ) \,.
	\end{equation}
An analogous expression for \ac{OFDM}, owing to the fact that the channel matrix in OFDM is diagonal, yields
	\begin{equation}
		C_\mathrm{Gauss}^{\mathrm{OFDM}}= \frac{T}{T+T_{\rm GI}} \log_2\left(1+\mathrm{SNR_{com}}\right).
	\end{equation}
The factors $\frac{NT}{NT + T_{\rm GI}}$ and $\frac{T}{T+T_{\rm GI}}$ for OTFS and OFDM, respectively, 
are introduced in order to take into account the insertion of the guard interval. In OTFS, a guard interval of duration $T_{\rm GI}$ is inserted at the end of
each frame of duration $NT$, comprising $N$ symbols in the time domain. 
In contrast, in OFDM, the guard interval in the form of cyclic prefix is inserted at each OFDM symbol of duration $T$.
It is clear that for practical values of the OTFS frame length $N$, the overhead paid by OTFS is much less than 
the CP overhead paid by OFDM (in  these results we used  $T_\mathrm{GI}=T/4$, which is typical in the IEEE 802.11 family of standards).
On the other hand, the larger overhead incurred by OFDM yields a particularly simple receiver structure, since the channel
matrix is diagonalized. In contrast, as we have seen extensively in this paper, OTFS requires block-wise detection
over the whole frame, which can be very computationally intensive. This is why the proposed soft-output symbol detector
presented in Section \ref{algo-G} is of particular interest, as it will be demonstrated by the results of the next subsection. 

\begin{figure}
		\centering
		\includegraphics[scale=0.8]{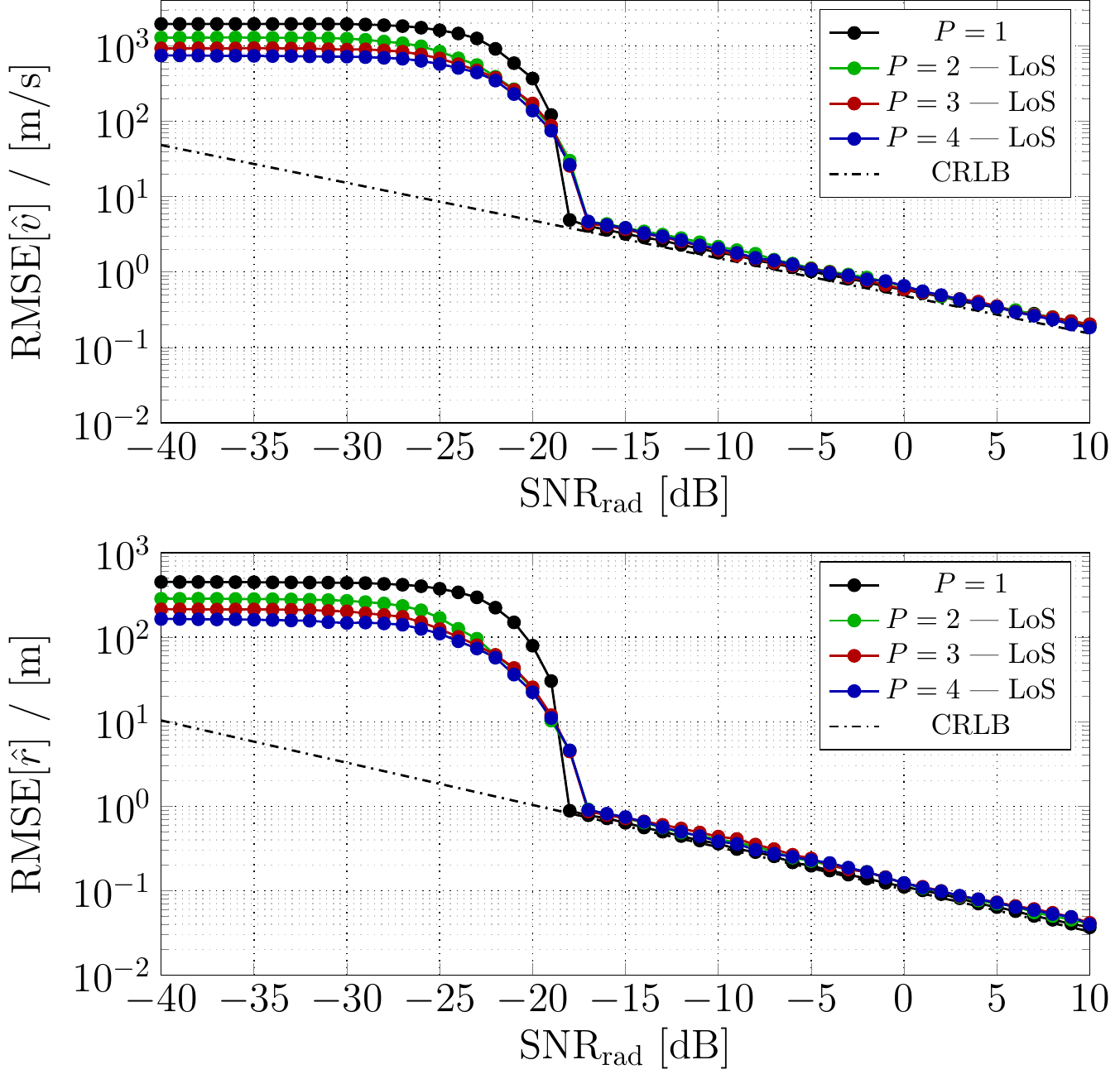}
		\caption{RMSE of the target velocity estimation $\hat{v}$ vs $\mathrm{SNR_{rad}}$ and range estimation $\hat{r}$ vs $\mathrm{SNR_{rad}}$ for a multi-path channel ($P\in\left[1,2,3,4\right]$).}
		\label{fig:MSE-multiple-paths}
	\end{figure}

Next we present a second set of results where we consider only \ac{OTFS} in the presence of  multipath channels ($P > 1$, up to $P=4$) 
and show the effectiveness of our proposed approximated \ac{ML} parameter estimator with at most 5 iterations, 
under the assumption that paths are enough spaced in the Doppler-delay grid. 
In our analysis, we suppose that the number of paths is known at the radar receiver. While this is not realistic, extensions to a threshold-based 
path detection approach with associated false alarm and missed detection probabilities goes beyond the scope of this paper and is deferred to future work. 
In Fig. \ref{fig:MSE-multiple-paths} we show range and velocity RMSE in a multipath scenario, where of course the parameter of interest are 
those of the \ac{LoS} path. As expected, the performance slightly degrades as the number of paths increases, but 
such degradation is very mild, showing the robustness of the proposed joint parameter estimation method. 
Notice also that the \ac{CRLB} is plotted for the case $P = 1$ only. 

	\subsection{Performance of Separated Detection and Decoding}

As already mentioned, we characterize the performance of separated detection and decoding schemes in terms of pragmatic capacity,  
i.e., the mutual information of the virtual channel with input the constellation symbols used with uniform probability and output 
provided by the soft-output of the detector. This mutual information provides an achievable rate for separated detection and decoding for a given detection 
scheme \cite{kavcic2003binary,soriaga2007determining}. 
Let us consider a sequence of $NM$ symbols $\left\{x_k\right\}$, belonging to the signal constellation $\mathcal{C}$, and let 
$V_k(x_k)$ denote the  detector soft-output. In the case of MP-based detectors, $V_k(x_k)$ is given in the form of a posterior probability distribution on
$x_k \in \mathcal{C}$, while in the case of linear equalizers (e.g., the linear MMSE estimator in (\ref{lmmse-eq})), this is given as the noisy estimate
$\hat{x}_k$ which is treated as the output of a (virtual) AWGN channel.  
In any case, the pragmatic capacity is simply defined as the symbol-by-symbol mutual information 
$I_\mathrm{PC}\left(x_k; V(x_k) \right)$.  When $V(x_k)$ takes on the form of a posterior probability distribution, this can be easily calculated 
by Monte Carlo simulation via the formula 
	\begin{align}\label{eq:Pragmatic-Capacity}
	I_\mathrm{PC}\left(x_k; V(x_k) \right) &\triangleq H(x_k) - H(x_k|V(x_k)) = \log_2\mathcal{C} - E\left\{ \sum_{x_k \in \mathcal{C}} V(x_k) \log_2\frac{1}{V(x_k)} \right\},  
	\end{align}
	where the expectation is obtained by Monte Carlo simulation from the output of the detector.  In the case of linear equalization (i.e.,  $V(x_k) = \hat{x}_k$), 
	the pragmatic capacity is simply given by the {\em symmetric capacity} (i.e., with symbols used with uniform probability)  of the signal constellation $\mathcal{C}$, for an AWGN channel with SNR equal to the output \ac{SINR} of the equalizer.
For the sake of comparison, we also show the symmetric capacity of the signal constellation in an AWGN channel with SNR equal to 
$\mathrm{SNR_{com}}$ (i.e., the SNR of the \ac{LoS} path), denoted by $C_\mathrm{AWGN}^\mathrm{sym}$, and the mutual information with
Gaussian inputs $C_\mathrm{Gauss}^\mathrm{OTFS}$, for the case $P = 1$.

\begin{figure}
		\centering
		\includegraphics[scale=0.72]{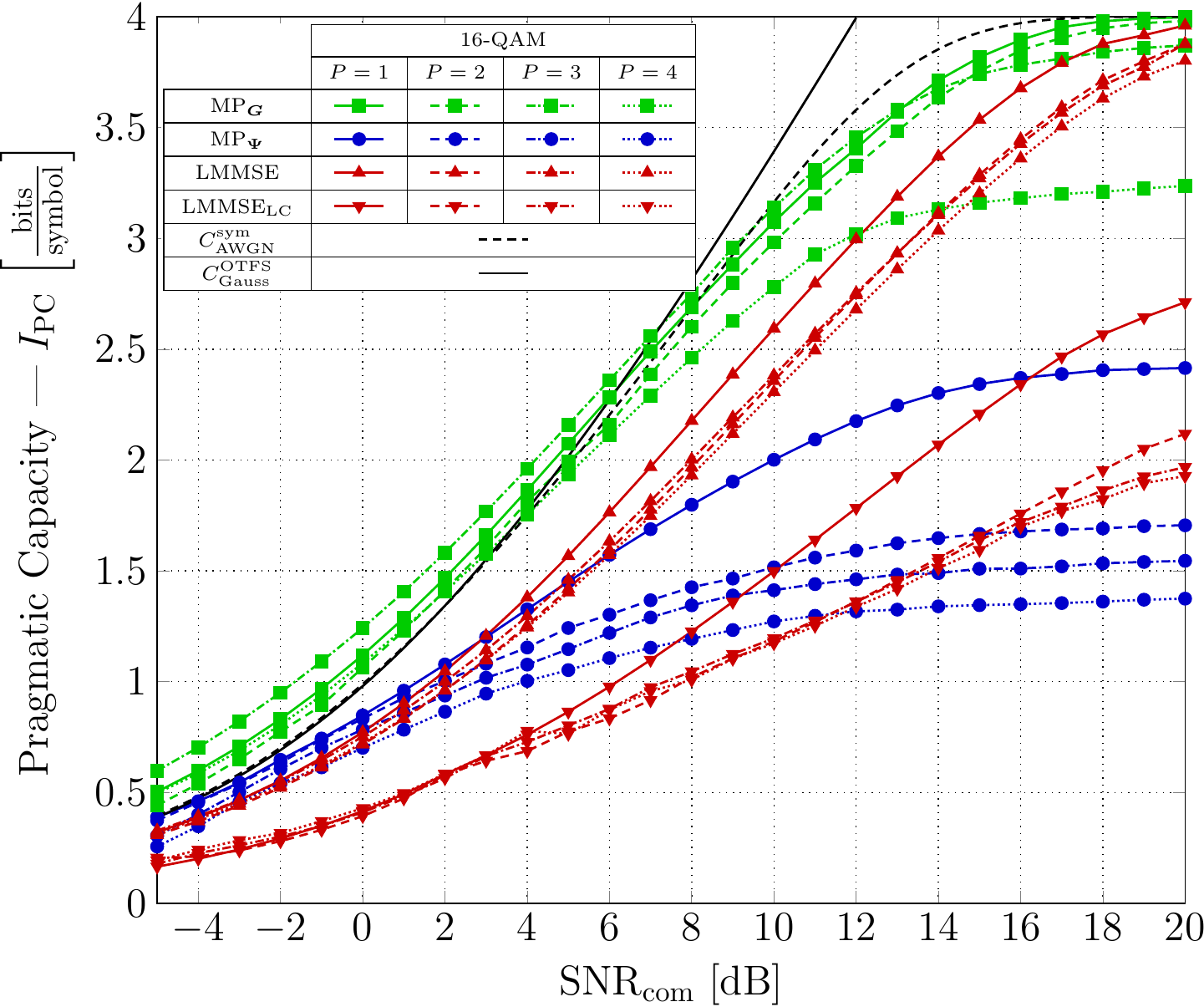}
		\caption{Symbols detection performance in terms of pragmatic capacity for 16-QAM modulation. The curves show the behavior of matrix $\boldsymbol{G}$ based \ac{MP} algorithm and \ac{MP} algorithm of \cite{raviteja2018interference} under approximated channel conditions, i.e., with delay and Doppler shifts on the Doppler-delay grid, for a multi-path channel with different number of components $P$.}
		\label{fig:Pragmatic-Capacity-HighMod-ApproxCh}
	\end{figure}

\begin{figure}
		\centering
		\includegraphics[scale=0.72]{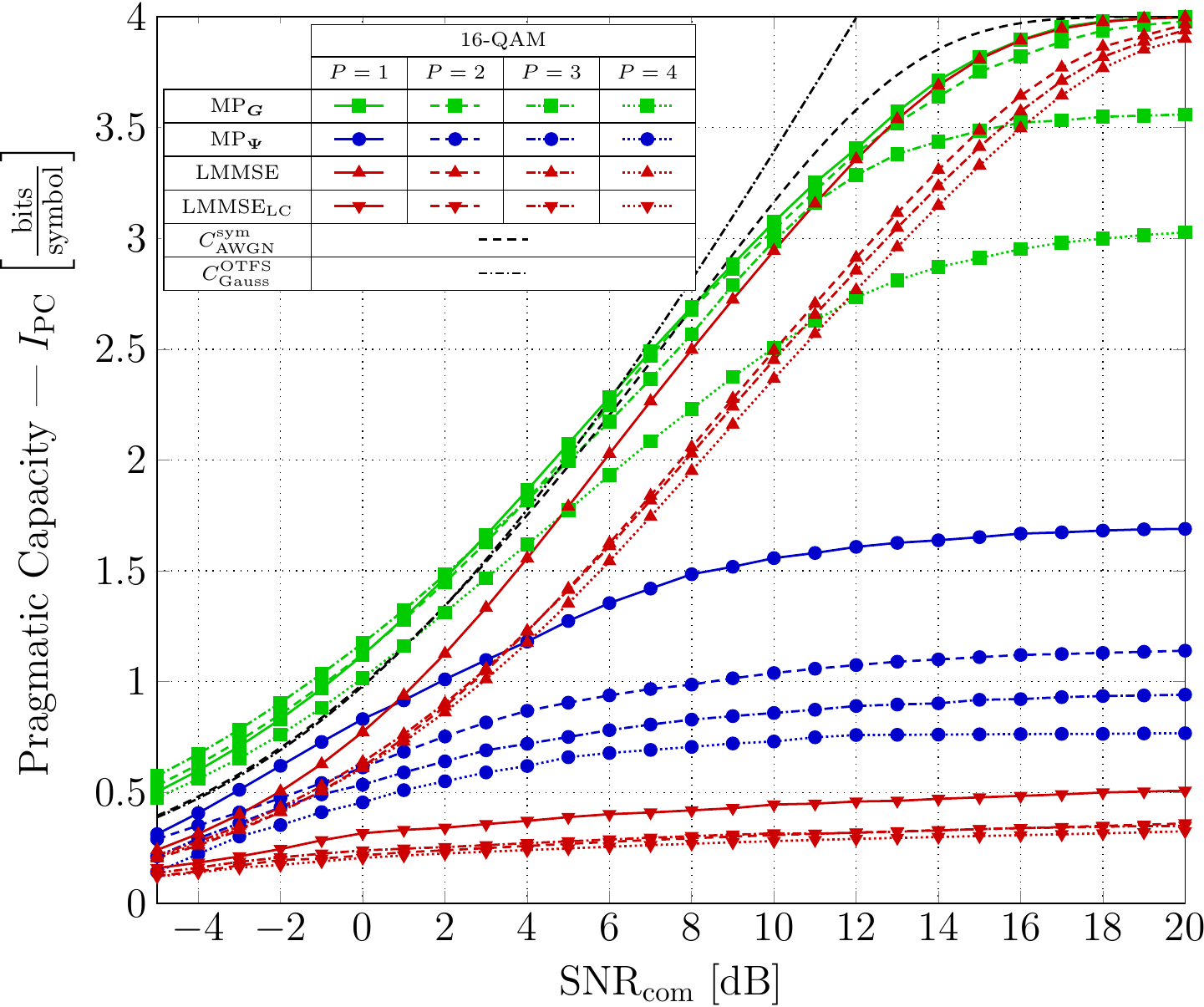}
		\caption{Symbols detection performance in terms of pragmatic capacity for 16-QAM modulation. The curves show the behavior of matrix $\boldsymbol{G}$ based \ac{MP} algorithm and \ac{MP} algorithm of \cite{raviteja2018interference} under real channel conditions, i.e., with delay and Doppler shifts not on the Doppler-delay grid, for a multi-path channel with different number of components $P$.}
		\label{fig:Pragmatic-Capacity-HighMod-RealCh}
	\end{figure}

	Fig. \ref{fig:Pragmatic-Capacity-HighMod-ApproxCh} and Fig. \ref{fig:Pragmatic-Capacity-HighMod-RealCh} show the 
	performance of the various methods for 
	\ac{OTFS} soft-output detection considered in Section \ref{sec:Detection-Algorithm} for a 16-\ac{QAM} modulation.
	Fig. \ref{fig:Pragmatic-Capacity-HighMod-ApproxCh} shows the results for the (unrealistic) case where
the channel Doppler shifts and delays are exactly on the discrete Doppler-delay grid used by the receiver sampling. 
In contrast, Fig. \ref{fig:Pragmatic-Capacity-HighMod-RealCh} shows the results when the actual channel has arbitrary Doppler and delay shifts 
(with a random uniformly distributed fractional part), but certain algorithms {\em assume} such integer grid when constructing the nominal channel matrix 
$\boldsymbol{\Psi}$ used by the detector (as advocated for example in  \cite{raviteja2018interference}. 

We notice that the proposed \ac{MP}-based approach outperforms the one in \cite{raviteja2018interference} in both cases. 
In particular, it suffers from almost no degradation due to the non-integer Doppler and delay shifts, unlike the method of
\cite{raviteja2018interference}.  
The \ac{LMMSE} equalizer (\ref{lmmse-eq}) with full complexity yields very good performance, paying only a small SNR penalty with respect to
the proposed MP-based scheme for $P = 1,2$, and outperforming the MP-based scheme for richer scattering $P = 3,4$. 
However, as said before, its complexity is cubic in the frame dimension $MN$, which is unaffordable in practical implementations. 
Unfortunately, the low-complexity \ac{LMMSE} estimator (curves indicated by $\mathrm{LMMSE}_{\mathrm{LC}}$ in the figures)
of \cite{cheng2019low} exploits a specific structure of the channel matrix $\boldsymbol{\Psi}$.
The required doubly block circulant feature of the \ac{OTFS} channel matrix, as defined in \cite{cheng2019low}, 
is satisfied only when $\gTX$ and $\gRX$ are bi-orthogonal. Under such condition, the scheme proposed in \cite{cheng2019low}, coincides with the 
MMSE block equalizer without the need of a large matrix inversion. 
In our model, we adopted physically realizable and realistic rectangular pulses, which clearly do not satisfy the bi-orthogonal condition. 
As a consequence, the doubly block circulant feature is lost. Our simulations show that the approach \cite{cheng2019low} is 
not competitive when applied to a channel model using rectangular pulses. It should be noticed that bi-orthogonality for pulses with time-frequency 
product equal to 1 is mathematically impossible \cite{matz2013time}. Hence, relaying on such assumption may be very misleading, as shown by our results.

\section{Conclusions}\label{sec:conclusions}

	We studied a joint radar and communication system based on the \ac{OTFS} modulation format, over a time-frequency selective channels formed by 
	discrete multipath components, each of which is characterized by a complex amplitude, a Doppler shift, and a delay. The radar parameters of interest
	are the delay and the Doppler shift of the shortest path (assumed to be in \ac{LoS} propagation), related to the range and velocity of the target. 
	We derived an efficient approximated ML parameter estimation scheme as well as bounds and tight approximations on the estimation MSE of the radar detector. The proposed estimation scheme for OTFS provides as accurate estimation as state-of-the art dedicated radar waveforms such as 
	\ac{FMCW}, while the digitally modulated signal is used to transmit information at full rate (i.e., all symbols are coded information symbols). 
	This show that joint radar estimation and data communication can be achieved with virtually no penalty for each of the functions, 
	at the only cost of complexity of the radar detector, which is significantly more complicated than the corresponding detector for \ac{FMCW}.
	
	We also considered a practical low-complexity soft-output detector for OTFS separated detection and decoding based on 
	message passing, derived by applying the canonical \ac{SPA} computation rules to a particular realization of the \ac{FG} of the underlying joint posterior probability distribution of the modulation symbols given the received signal frame.  
	The proposed MP-based soft-output detector shows very good performance at low complexity (especially because it is suitable for a highly parallelized implementation). In particular, the proposed scheme outperforms other recently proposed MP-based schemes, and 
	provides a competitive performance/complexity tradeoff with respect to the block-wise linear MMSE detector, which 
	is prohibitively complex.  As such, the proposed detector represents the new state-of-the-art for OTFS soft-output detection.
	An interesting issue for future work is to investigate the performance of ``Turbo Equalization'', when the soft-output detector is iteratively re-processed
	using the soft-ouptut of an outer channel decoder. Our proposed scheme appears to be ideally suited and easily adapted to
	such Turbo Equalization schemes. 
	
	\section{Acknowledgment}
	
	The work of Lorenzo Gaudio, Giuseppe Caire, and Giulio Colavolpe is supported by Fondazione Cariparma, under the TeachInParma Project. The work of Mari Kobayashi is supported by an Alexander von Humboldt Research Fellowship. The authors would like to thank Bj\"orn Bissinger for his support on an early version of this paper.

	\bibliography{IEEEabrv,tradeoff}
	
\end{document}